\documentclass[a4paper, 10pt,conference]{ieeeconf}      % Use this line for a4
\IEEEoverridecommandlockouts                              % This command is only
\makeatletter
\let\NAT@parse\undefined
\makeatother

\usepackage[dvipsnames]{xcolor}

\newcommand*\linkcolours{ForestGreen}

\usepackage{times}
\usepackage{graphicx}
\usepackage{subcaption}
\usepackage{amssymb}
\usepackage{siunitx}
\usepackage{gensymb}
\usepackage[export]{adjustbox}
\usepackage{amsmath}
\usepackage{breakurl}

\usepackage{url,hyperref}
\hypersetup{
colorlinks,
linkcolor=\linkcolours,
citecolor=\linkcolours,
filecolor=\linkcolours,
urlcolor=\linkcolours}

\usepackage{algorithm}
\usepackage{algorithmic}

\usepackage[labelfont={bf},font=small]{caption}
\usepackage[none]{hyphenat}

\usepackage{mathtools, cuted}

\usepackage[noadjust, nobreak]{cite}
\usepackage{tabularx}
\usepackage{amsmath}

\usepackage{float}

\PassOptionsToPackage{hyphens}{url}\usepackage{hyperref}
\usepackage{pifont}% http://ctan.org/pkg/pifont

\newcolumntype{Y}{>{\centering\arraybackslash}X}

\usepackage[]{placeins}

\usepackage{placeins}

\usepackage{tikz}

\usepackage[framemethod=tikz]{mdframed}

\usepackage{afterpage}

\usepackage{stfloats}

\usepackage{atbegshi}
\newcommand{\handlethispage}{}
\newcommand{\discardpagesfromhere}{\let\handlethispage\AtBeginShipoutDiscard}
\newcommand{\keeppagesfromhere}{\let\handlethispage\relax}
\AtBeginShipout{\handlethispage}

\usepackage{comment}

\title{\LARGE \bf
Synergy between Observation Systems Oceanic in Turbulent Regions
}

\author{Van-Khoa Nguyen, Santiago Agudelo\\
IMT Atlantique, Computer Science Department, CS 83818 F-29238, Cedex 3, Brest, France\\
{\tt\small 1411836@hcmut.edu.vn,santiago.agudelo-ramirez@imt-atlantique.net}
}

\begin{document}
\maketitle
\pagestyle{plain}
%=======================================================================================================
\begin{abstract}

Ocean dynamics constitute a source of incertitude in determining the ocean's role in complex climatic phenomena. Current observation systems have limitations in achieving sufficiently statistical precision for three-dimensional oceanic data. It is crucial knowledge to describe the behavior of internal ocean structures.  We present the data-driven approaches which explore latent class regressions and deep regression neural networks in modeling ocean dynamics in the extensions of Gulf Stream and Kuroshio currents. The obtained results show  a promising data-driven direction for understanding the ocean's characteristics, including salinity and temperature, in both spatial and temporal dimensions in the turbulent regions. Our source codes are publicly available at \href{https://github.com/v18nguye/gulfstream-lrm}{https://github.com/v18nguye/gulfstream-lrm} and at \href{https://github.com/sagudelor/Kuroshio-Turbulent-Region-Analysis-by-Neural-Network/tree/master}{https://github.com/sagudelor/Kuroshio}\url{}.

\end{abstract}

\section{INTRODUCTION}
The ocean is a dynamic, turbulent and chaotic system, difficult to understand and predict. This is particularly true in western oceanic regions of medium latitudes, where significant currents shrink into intense fronts. Such regions are called "western border extension currents". The turbulence of these regions constitute one of the main incertitude sources in evaluating the role of the ocean in climate. Such incertitude can be lowered if we ameliorate our capacity of diagnostic of the temporal evolution of such oceanic structures.\\

None of the current oceanic observation systems can provide with sufficient precision a time series of thermal three dimensional fields for these problematic currents. Satellite measurements have a high local frequency and a precise horizontal resolution, but they only manage to capture the surface signature of intern oceanic structures. Databases like Argo which perform in-situ measurements of oceanic variables provide high precision vertical information, but they lack of horizontal resolution. \\

The objective of this paper is to describe two different approaches to model ocean's dynamics in two different problematic regions. The first approach uses a latent class regression model to discover underlying distribution shapes of certain ocean's characteristics at the extension of Gulf Stream. Such method considers the presence of different dynamic modes that can be observed in turbulent regions, each of which is represented by a local regression. The second approach uses neural networks to predict ocean's salinity and temperature profiles from  available data at the extension of Kuroshio's current.

\subsection*{Argo database}
Argo is an array of 3000 free-drifting floats that measure the temperature and salinity of the upper 2000m of the ocean. Information provided by Argo allows monitoring temperature, salinity, velocity and, in some cases, bio-geo parameters such as oxygen or chlorophyll concentration.
Data collected by Argo becomes publicly available in near-real time through the Global Data Assembly Centers in Brest and Monterey after automatic quality control \cite{UCSD2020}.

\subsection*{Copernicus Marine environment monitoring service}
The CMEMS  provides core reference information on the state of the physical oceans and regional seas. Different products from CMEMS were used for validation purposes, more specifically to extract data concerning sea surface temperature (SST) and sea surface height (SSH) \cite{CMEMS2020}, \cite{CMEMS2020a}, \cite{CMEMS2020b}.

\section{METHODOLOGY}

\subsection{First approach : latent class regression model}
\subsubsection{Latent class regression model}
 In the statistical domain, the finite mixture model has proven its efficiency, been widely used in data analysis \cite{finteGeoffrey}. In this paper, we focus on one of its methods, which is  the unsupervised latent class regression method, to discover underlying distribution shapes of certain  ocean’s characteristics. An ocean’s characteristic, which we call a target $Y$, is modeled by summing $K$ linear regression models on its feature set $X$. Each linear regression is interpreted as a dynamical mode, which characterizes and varies at different local regions in the ocean. Our objective is to identify an optimal value for $K$ number of dynamical modes, in order to approximately represent structure ocean’s characteristics, as well as to avoid the over-fitting effect. In the following analysis,  we assume that the conditional probability of a target $Y$ given its feature set X and a dynamical mode  $Z = k$ is given as a multivariate Gaussian distribution, which mean and co-variance respectively correspond to $X\beta_{k}$ and $\Sigma_{k}$:

\begin{equation}
P(Y|X,Z=k) \propto \mathcal{N}_{k}(Y;X\beta_{k},\Sigma_{k}).
\end{equation}

The advantage of the latent regression method is to efficiently decompose a global distribution into separate local ones. By considering $K$ dynamical modes, the conditional probability of  $Y$ given $X$ resorts to a mixture of Normal distribution:

\begin{equation}
P(Y|X) = \sum_{k=1}^K \lambda_{k}\mathcal{N}_{k}(Y;X\beta_{k},\Sigma_{k}).
\end{equation}

Each dynamical mode is assigned by a prior probability $\lambda_{k}$ representing its known individual contribution. By construction, it imposes constraints on the prior knowledge $0\leq \lambda_{k}\leq 1 $ and $ \sum_{k = 1}^K \lambda_{k}=1$. To simplify the notations, we summarize all model’s parameters as $\theta = (\lambda_{1},..,\lambda_{K},\beta_{1},...,\beta_{K},\Sigma_{1}...\Sigma_{K})$. In order to choose the model’s parameters which generate a maximum likelihood estimation, the iterative Expectation-Maximization (EM) scheme is given below.

\subsubsection{Model Training}
The chosen model should generate a better likelihood value in all possible parameter configuration. In \cite{6605600}, the authors introduced the EM procedure, which is used to find appropriate parameter’s values to maximize a the given log-likelihood function below:

\begin{equation}
\mathcal{L}(\theta) = \sum_{k=1}^n \log p(Y(i)|X(i),\theta).
\end{equation}

The EM algorithm is easy to program and proceeds iteratively in an E-step (Expectation-step) and M-step (Maximization-step). In the E-step, the objective is to estimate the posterior probability on the latent variable Z at each data point. The estimate for the $i^{th}$ data point in the $k^{th}$ dynamical mode, with current parameter value $\hat{\theta}$ given as below: 

\begin{equation}
\hat{\pi}_{k}(i) = \frac{\hat{\lambda}_{k}\mathcal{N}_{k}(Y(i),X(i)\hat{\beta}_{k},\hat{\Sigma}_{k})}{P(Y(i)|X(i),\hat\theta)} \forall k \in K
\end{equation}

The M-step update includes minimizing the expectation of the log-likelihood conditionally to the current parameter estimate $\theta$, and maximizing the model’s log-likelihood. The minimizing problem leads to update the prior and regression parameters, which are given below:

\begin{equation}
\hat{\pi}_{k}(i) = \frac{\sum_{i =0}^n \hat{\pi}_{k}(i)}{n} \forall k \in K
\end{equation}

\begin{equation}
\hat\beta{k} = (X^TW_{k}X)^{-1}X^TW_{k}Y
\end{equation}

Where $X$ is a matrix with a size n x number of features, $W_{k}$ is an nxn diagonal matrix with diagonal entries $\hat{\pi}_{k}(i)$, and $Y$ is a nx1 vector of observation.

The maximization of the model’s log-likelihood is derived from the derivative the log-likelihood function with respect to $\Sigma_{K}$, the parameter update is given as following:

\begin{equation}
\hat{\Sigma}_{k} = \frac{\sum_{1}^n\hat{\pi}_{k}(i)\epsilon_{k}(i)^T\epsilon_{k}(i)}{\sum_{1}^n\hat{\pi}_{k}(i)}, \forall k
\end{equation}

where $\epsilon_{k}(i) = Y(i) - X(i)\hat{\beta}_{k}$. The  algorithm iterates the E-step and the M-step until the upward variation of the log-likelihood is negligible to its current value.

The EM algorithm is sensitive to local maximization, to which we need an efficient initialization method. The K-means algorithm is used to cluster data points into K different dynamical modes, and uses it as an initialization step. The critical part of the EM algorithm is to choose an appropriate K number of dynamical modes, which satisfies the trade-off between the likelihood maximization and the model complexity. In statistical learning, one of the common criteria for evaluating the performance of the model is the Bayesian Information Criterion (BIC), which shows how the model is performing, where is the optimal number of parameters added to the model. The general principle of the BIC described as following:

\begin{equation}
BIC(K) = -2\mathcal{L}(K) + \mathcal{N}(K)\log{n}
\end{equation}

where $\mathcal{L}(K)$ is the log likelihood of trained model with K classes, $\mathcal{N}{k}$ is the number of parameters in the model, and n is the number of training data. We tend to choose the model which has a smaller BIC’s value than others. The BICs are penalized more if more parameters are added to the model to avoid the overfitting problem.

\subsection{Second approah : learning ocean's dynamics by using Neural Networks}
In this approach, data from  the Kuroshio current extension between years 1999 and 2016 were used. It is important to recall that availability from data has evolved within time. Indeed, technology improvements in recent years have made data collection much easier, resulting into more available data. \\

Pre-treatment of the geographical coordinates was necessary in order to get a projection allowing a cartesian representation of the zone we're interested in. This is why the neural network was fed with the cosine of the longitude and the sine of the latitude and not with the longitude and the latitude themselves. Other variables used as input data were the ocean's temperature and salinity (which are the variables we aim predicting with our model), the barometric pressure (which can be directly related to the depth at which the measure is taken), the sea level anomaly (sla) and the time at which measurements are available. In order to introduce a seasonality notion in our model, we  introduce the terms $sin(2\pi f t)$ and $cos(2\pi f t)$ where $f=1/(365 days)$ and $t$ is the time in days ; this way, we can keep record of the measurement instants corresponding to the same period of the year in different years.\\

Input data were normalized and fed into different neural networks which aim describing ocean's dynamics in terms of its temperature and salinity. Apart from the network's architecture, results may be sensible to different hyper-parameters such as batch size, optimization criteria,  optimizer and  activation function. \\

Optimization criteria we're willing to minimize is the mean square error between the predicted temperature and salinity and the measured ones, and to do so we've chosen Adam's optimizer. We also set relu as the network's activation function. \\

\begin{figure}
 \centering
 \begin{subfigure}[b]{0.2\textwidth}
     \raggedright
     \includegraphics[width=1.1\textwidth]{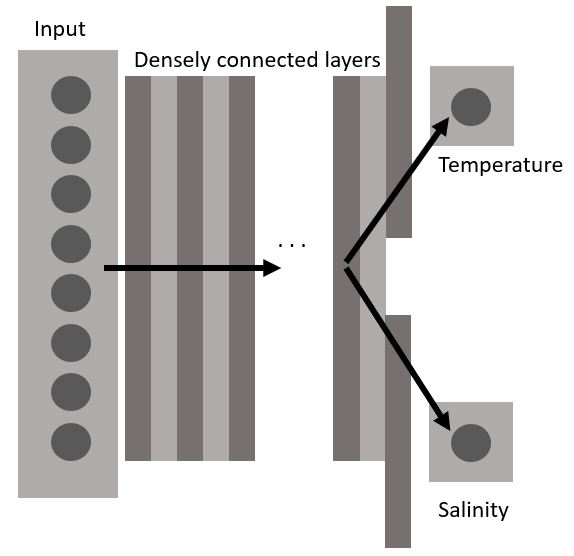}
     \caption{}
 \end{subfigure}
\hspace{1em}%
 \begin{subfigure}[b]{0.2\textwidth}
     \raggedleft
     \includegraphics[width=1.1\textwidth]{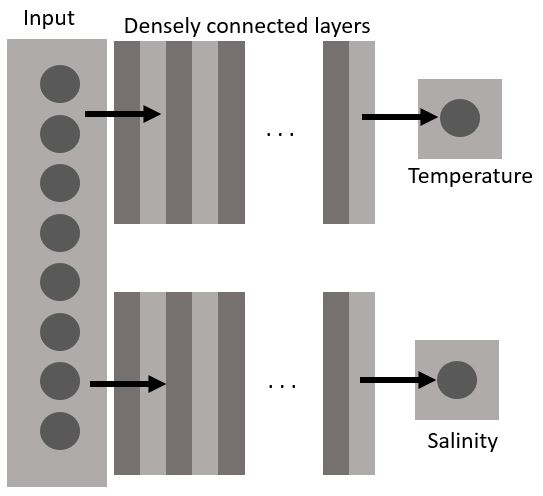}
     \caption{}
 \end{subfigure}
 \begin{subfigure}[b]{0.2\textwidth}
     \raggedright
     \includegraphics[width=1.1\textwidth]{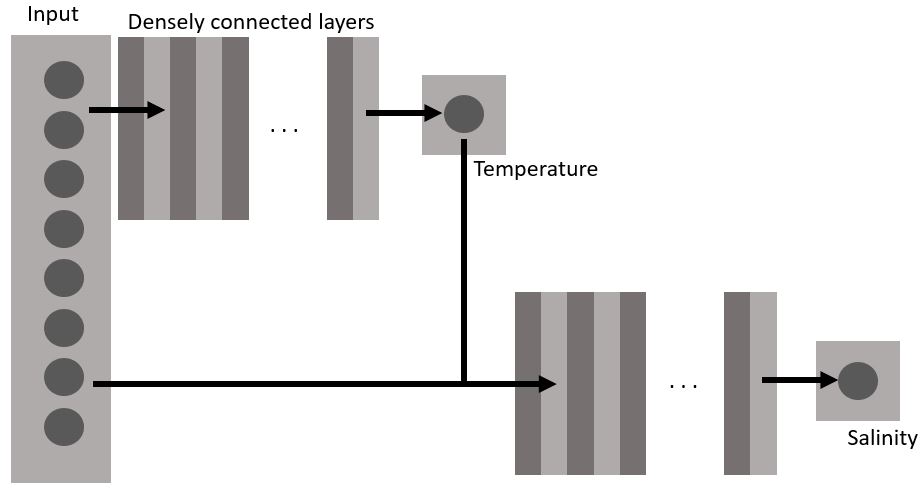}
     \caption{}
 \end{subfigure}
\hspace{1em}%
 \begin{subfigure}[b]{0.2\textwidth}
     \raggedleft
     \includegraphics[width=1.1\textwidth]{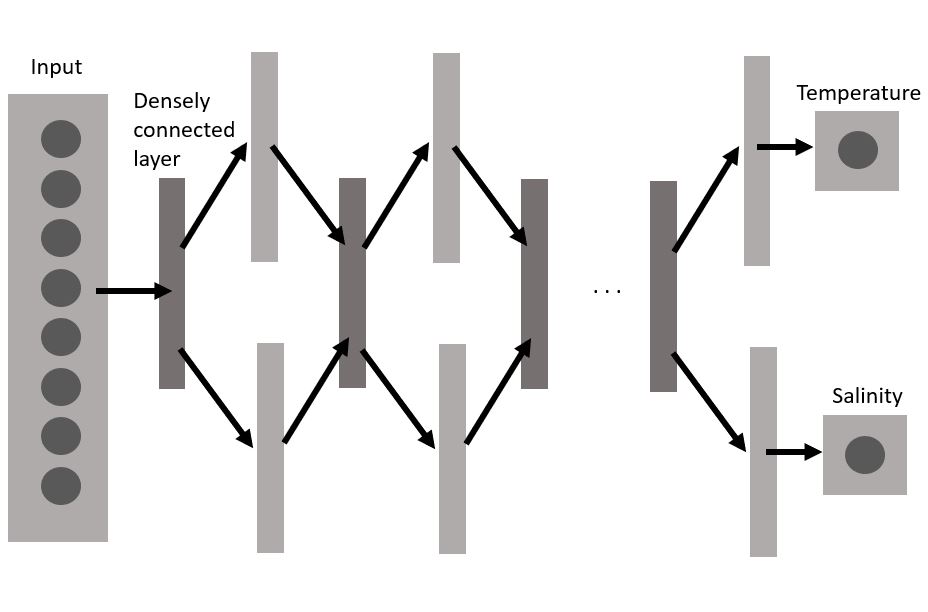}
     \caption{}
 \end{subfigure}
    \caption{Schemes of the different tested architectures (simple, parallel, cascade and junction, respectively) }
    \label{architecture}
\end{figure}

We tested four different architectures : 
\begin{itemize}
    \item Simple architecture: data are fed to a network consisting in several perceptron layers. Two outputs are obtained corresponding to  temperature and salinity of the ocean. 
    \item Parallel architecture: In this case, data are fed to two different networks, each of them consisting in several perceptron layers like in the "simple" architecture above. Each parallel network is used to predict one of the variables we're interested in (salinity or temperature). 
    \item Cascade architecture: Data are fed into a first network which allows predicting the temperature. This result as well as the input data are injected to a second network which is used to predict the salinity of the ocean. 
    \item Junction architecture: the input is fed into a layer which propagates data to two separate independent layers. The outputs of such layers are fed into a single layer, and the procedure is repeated as shown in Figure \ref{architecture} (d).
    
\end{itemize}

\begin{figure}
    \centering
    \includegraphics[width=0.5\textwidth]{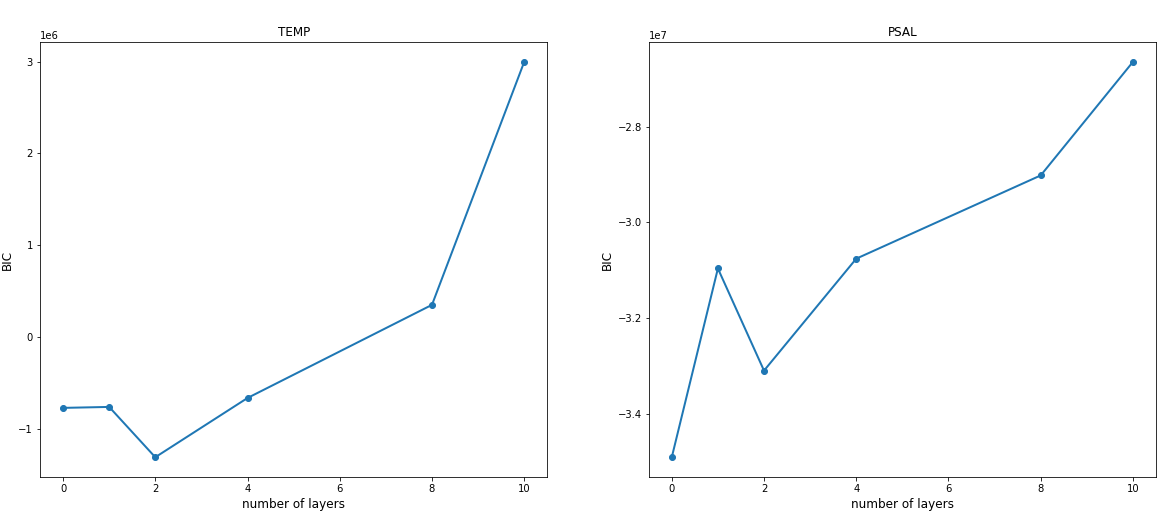}
    \caption{BIC in function of additional 128 unit layers added to each network in the parallel model}
    \label{bicparallel}
\end{figure}

The first layer of each network is composed of 128 units and the final layers are composed of 32 units each. For all the architectures, the total number of intermediate layers was varied in order to determine the influence of the model's complexity in the networks performance. Such analysis takes into account the temperature and salinity MSE for validation data in function of the number of layers added to the model as well as the model's complexity; to do so, the BIC indicator was calculated. Since adding layers (i.e, more parameters) may have different effects on temperature and salinity, the number of layers was defined as the one that best minimized both variables' associated BIC in average terms. For example, figure \ref{bicparallel} shows the computed BIC in function of the additional 128 unit layers added to each of the networks in the parallel architecture model. When adding 2 additional layers, the best trade-off between model's complexity and error minimization is reached. The Table \ref{table1} describes the composition of the networks that gave the best results for each of the considered architectures. \\

\begin{table}[t]
\centering
\resizebox{0.5\textwidth}{!}{\begin{tabular}{|l|l|l|l|}
\hline
\multicolumn{1}{|c|}{Architecture} & MSE TEMP  & MSE PSAL  & Description of the architecture                                                                                                                                                                   \\ \hline
Simple                             & 0.8312 & 0.0433 & \begin{tabular}[c]{@{}l@{}}7  layers of 128 units + two \\ layers of 32 units at the end of\\ the network\end{tabular}                                                                            \\ \hline
Parallel                           & 0.8073 & 0.0568 & \begin{tabular}[c]{@{}l@{}}3 layers of 128 units in each parallel \\ network + one layer of 32 units at\\ the end of each network.\end{tabular}                                                   \\ \hline
Cascade                            & 0.6964 & 0.0533 & \begin{tabular}[c]{@{}l@{}}5 layers of 128 units in the first \\ network + 4 layers of 128 units in the second \\ network + one layer of 32 units at the end of\\ each network.\end{tabular}      \\ \hline
Junction                           & 0.8177 & 0.0464 & \begin{tabular}[c]{@{}l@{}}Initial layer of 128 units + two times \\ (two separate layers  of 128 units + \\ intermediate layer of 64 units) + \\ two final layers of 32 units each.\end{tabular} \\ \hline
\end{tabular}}
\caption{Distribution of layers that gave the best performance for each of the considered architectures}
\label{table1}
\end{table}

In order to take into account possible temporal correlation between input data, we decided to test the influence of adding an LSTM layer  to the model. First, we took a simpler model consisting of a single 128 unit layer plus two 32 unit layers at the end. Then, we added an LSTM layer to this model and we determined the influence that the number of units on this layer had on the model's BIC indicator. We took the number of units minimizing this quantity and we added an LSTM layer with such number of units (50 in total) to the "simple" architecture model defined in Table \ref{table1} so we could be able to tell how this layer could improve its performance. This experiment was only performed on the "simple" architecture. \\

To avoid overfitting, the early stop method available on Tensorflow library was used. This prevents the validation error to increase and so the retained model fits the validation data as well as it fits the training data. 

%=======================================================================================================
\section{Results}
\subsection{First approach' results}
\subsubsection{Training process}
\begin{figure}
 \centering
 \begin{subfigure}[b]{0.2\textwidth}
     \raggedright
     \includegraphics[width=1.2\textwidth]{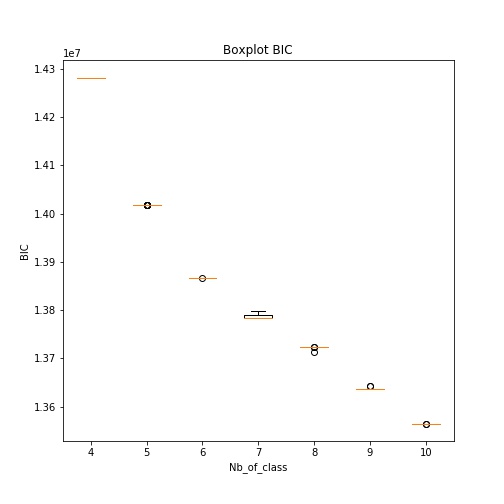}
     \caption{}
 \end{subfigure}
\hspace{1em}%
 \begin{subfigure}[b]{0.2\textwidth}
     \raggedleft
     \includegraphics[width=1.2\textwidth]{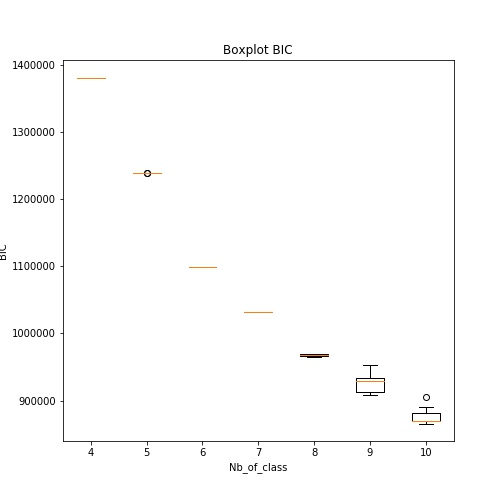}
     \caption{}
 \end{subfigure}
    \caption{ BIC box plots for the different temperature (a) and  salinity (b) models}
    \label{boxplot}
\end{figure}

In the training step, we ran  two independent models for temperature and salinity. For each model, we tested a range value from 4 to 10 number dynamical modes, each dynamical mode was initialized 10 times to get a consistency result. The models were trained on the Google Colab server, and it took 12 hours of training for each model. The  fig.~\ref{boxplot} depicts the box plots of BIC’s values  in the function of the number of the dynamical modes, from which we deduct the number of model's parameters: $\mathcal{N}(K) = K + K*T*T + K*D*T$ where K is the number of dynamical modes, T is the number predicted targets and D is the number of data's features.

As we can see when adding more dynamical modes the model gets  lower  BIC values until they reach their minimum. Theoretically, we should choose the model which yields the lowest BIC value. But in terms of result interpretation in oceanography, too many dynamical modes can make the results becoming hardly interpretable as we could not recognize the main characteristic of each mode.  To address the problem, we used the elbow heuristic method, and chose K = 8 as the appropriate number of dynamical modes for both models.\\

\subsubsection{Vertical and horizontal ocean’s characteristic interpolation}

\begin{figure}
 \centering
 \begin{subfigure}[b]{0.2\textwidth}
     \raggedright
     \includegraphics[width=1.1\textwidth]{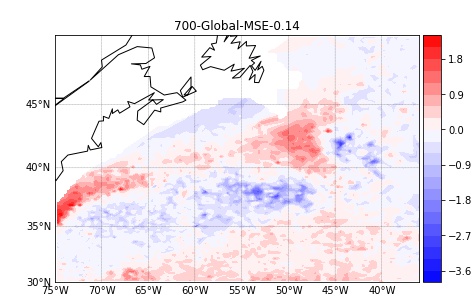}
     \caption{}
 \end{subfigure}
\hspace{1em}%
 \begin{subfigure}[b]{0.2\textwidth}
     \raggedleft
     \includegraphics[width=1.1\textwidth]{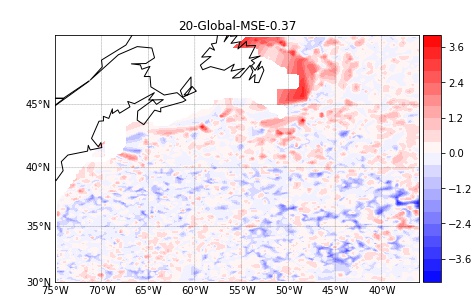}
     \caption{}
 \end{subfigure}
 \begin{subfigure}[b]{0.2\textwidth}
     \raggedright
     \includegraphics[width=1.1\textwidth]{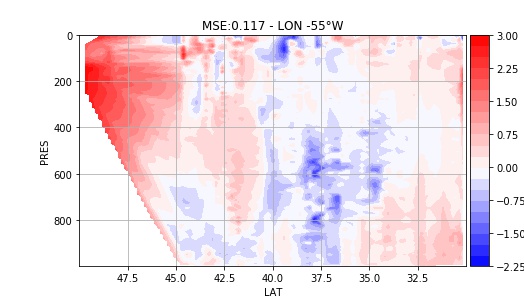}
     \caption{}
 \end{subfigure}
\hspace{1em}%
 \begin{subfigure}[b]{0.2\textwidth}
     \raggedleft
     \includegraphics[width=1.1\textwidth]{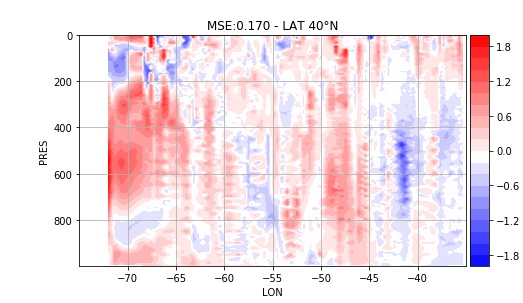}
     \caption{}
 \end{subfigure}
    \caption{Temperature interpolation’ error through a cut (a) at the pressure 700 - mse = 0.14\si{\degree}C, (b) at the pressure 20 - mse = 0.37\si{\degree}C, (c) at the longitude 55\si{\degree}W, - mse = 0.117\si{\degree}C (d) at the latitude 40\si{\degree}N - mse = 0.170\si{\degree}C }
    \label{erro-interpolation}
\end{figure}

Understanding the vertical and horizontal distribution of the ocean’s characteristics is a crucial role to further analyses in oceanography. By collecting certain oceanic information, which is described in the data section,  manipulating it, and treating it as features. The trained model should have the capacity of interpolating salinity and temperature at any local space in the ocean. In terms of climatology, we are going to globally analyse how the model can interpolate salinity and temperature distribution in all collected historic data. The fig.~\ref{erro-interpolation} shows temperature interpolation’s error  though three dimensional ocean’s space.

\begin{figure}
 \centering
 \begin{subfigure}[b]{0.2\textwidth}
     \raggedright
     \includegraphics[width=1.1\textwidth]{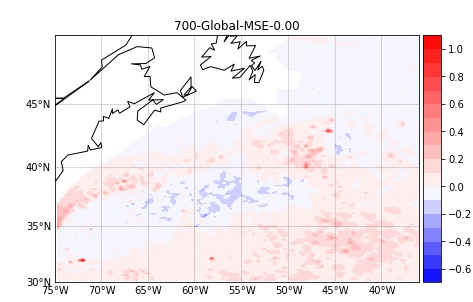}
     \caption{}
 \end{subfigure}
\hspace{1em}%
 \begin{subfigure}[b]{0.2\textwidth}
     \raggedleft
     \includegraphics[width=1.1\textwidth]{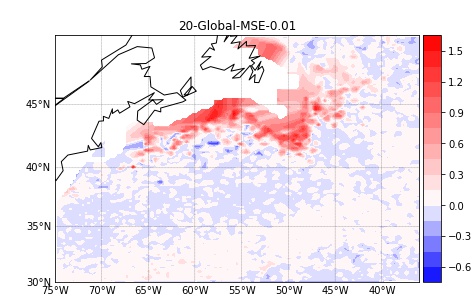}
     \caption{}
 \end{subfigure}
 \begin{subfigure}[b]{0.2\textwidth}
     \raggedright
     \includegraphics[width=1.1\textwidth]{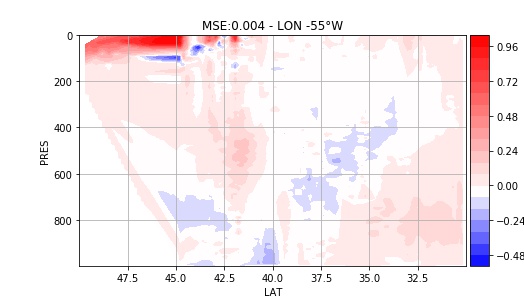}
     \caption{}
 \end{subfigure}
\hspace{1em}%
 \begin{subfigure}[b]{0.2\textwidth}
     \raggedleft
     \includegraphics[width=1.1\textwidth]{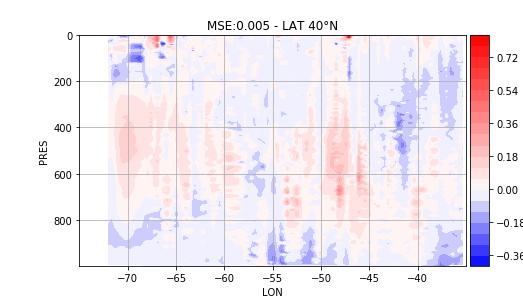}
     \caption{}
 \end{subfigure}
    \caption{Salinity interpolation’ error through a cut (a) at the pressure 700 - mse = 0.000psu, (b) at the pressure 20 - mse = 0.010psu, (c) at the longitude 55\si{\degree}W, - mse = 0.004psu (d) at the latitude 40\si{\degree}N - mse = 0.005psu }
    \label{sal-error-interpolation}
\end{figure}

Certain regions in these cuts have significantly higher errors than the rest of the cuts, as almost in these regions the data collection was interrupted and discontinuous in the time. The mean squared error statistic was used to evaluate the interpolating model, some cuts show significantly lower error in terms of climatology. The same analysis was made for salinity, the figure fig.~\ref{sal-error-interpolation} shows salinity interpolation’s error through the same cuts as the precedent case.\\

The global salinity’s ocean change evolves slowly in time, its values range from 34 PSU to  37 PSU. The fact of precise salinity interpolation in the ocean is therefore very important to understand its evolution. The four cuts through the ocean yielded quite precise interpolation results for salinity. For more details about the ground truth and estimated distributions of salinity and temperature of the precedent cuts, please reference to the fig.~\ref{gt-est} in the Appendix.

\subsubsection{Seasonal variations in seawater temperature and salinity}
In physical oceanography, seawater temperature and salinity have its own seasonal variations. At shallow water regions, these seasonal characteristics tend to be more clear compared with depth water regions. To estimate the real seasonal variations at a specific local region in the ocean, we estimated each daily salinity, temperature  for a temporal series in several consecutive years. For each instant, to predict the seawater temperature and salinity with a given feature, we needed information relating to prior mode probabilities. In order to possess this type of information, we used available prior probabilities calculated in the training step with historically recorded data and made an interpolation based on the training prior mode probabilities in that region. When the prior had been estimated for each instant in the temporal series, we used the fuzzy regression to estimate the seawater temperature and salinity:

\begin{equation}
\hat{Y}(i) = \sum_{k=0}^{K}\hat{\pi}_{k}(i)X(i)\hat{\mathcal{B}}_{k}
\end{equation}

We calculated then a mean value for each day in a year from the estimated consecutive year's seawater temperature and salinity respectively. The mean variation in a year is principally made up of a trend component $\tau_{t}$, a cyclical components $c_{t}$, and a random component as error $e_{t}$.  To get its trend component, we proposed to use then the Hodrick-Prescott filter to decompose the mean variations in a year. The Hodrick-Prescott principle \cite{Hodrick-Prescott} is to find a trend component in a temporal series $y_{t}$, which will solves:

\begin{equation}
\min_{\tau}(\sum_{t = 1}^T(y_{t}-\tau_{t})^2 + \lambda\sum_{t = 2}^{T-1}[(\tau_{t+1}-\tau_{t}) - (\tau_{t}-\tau_{t-1})]^2)
\end{equation}

\begin{figure}
 \centering
 \begin{subfigure}[b]{0.4\textwidth}
     \raggedright
     \includegraphics[width=\textwidth]{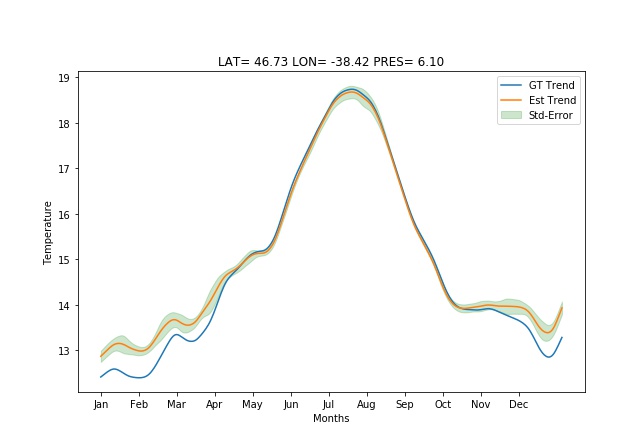}
     \caption{}
 \end{subfigure}
\hspace{1em}%
 \begin{subfigure}[b]{0.4\textwidth}
     \raggedleft
     \includegraphics[width=\textwidth]{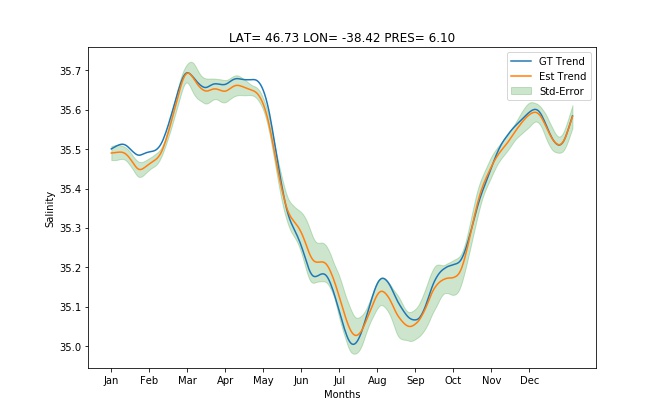}
     \caption{}
 \end{subfigure}
    \caption{ Seasonal variations at the surface in seawater (a) temperature and (b) salinity}
    \label{seasonal}
\end{figure}

The $\lambda$ term is used to penalize the grow rate of the trend component, and chosen as 1000 in our experiences. The fig.~\ref{seasonal} shows the ground truths and estimations of seawater temperature and salinity in the local ocean’s region which coordinates are 46.73\si{\degree}N in latitude, 38.42\si{\degree}W in longitude and 6.1 in pressure.

The estimated results reflected well the ground truth seawater temperature and salinity variations and had approximate amplitude’s values.  The temperature shows maxima in July and early Autumn and minimum in winter and early spring at the surface. Inversely to seawater temperature cycle, the salinity attains its maxima in March and April and its minimum in July and August. As we can see a negative correlation between seawater temperature and salinity, which is truly in the physical oceanography phenomenon. A precise prediction of the variations could help to correctly interpolate other ocean phenomena.\\

\subsubsection{Characterization of ocean dynamics}
\begin{figure}
\raggedright
 \begin{subfigure}[b]{0.5\textwidth}
     \centering
     \includegraphics[width=\textwidth]{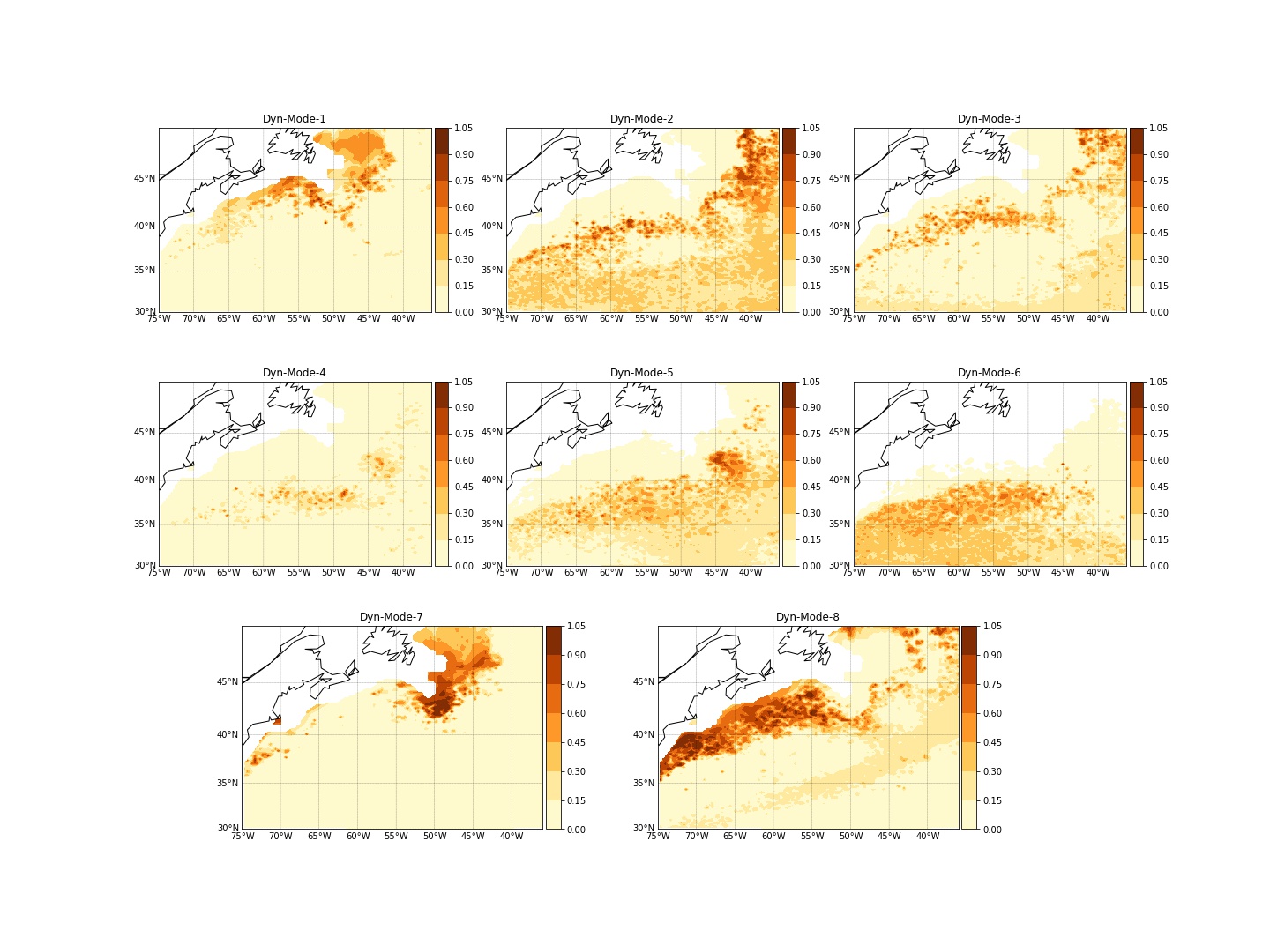}
     \caption{}
 \end{subfigure}
 \begin{subfigure}[b]{0.5\textwidth}
     \centering
     \includegraphics[width=\textwidth]{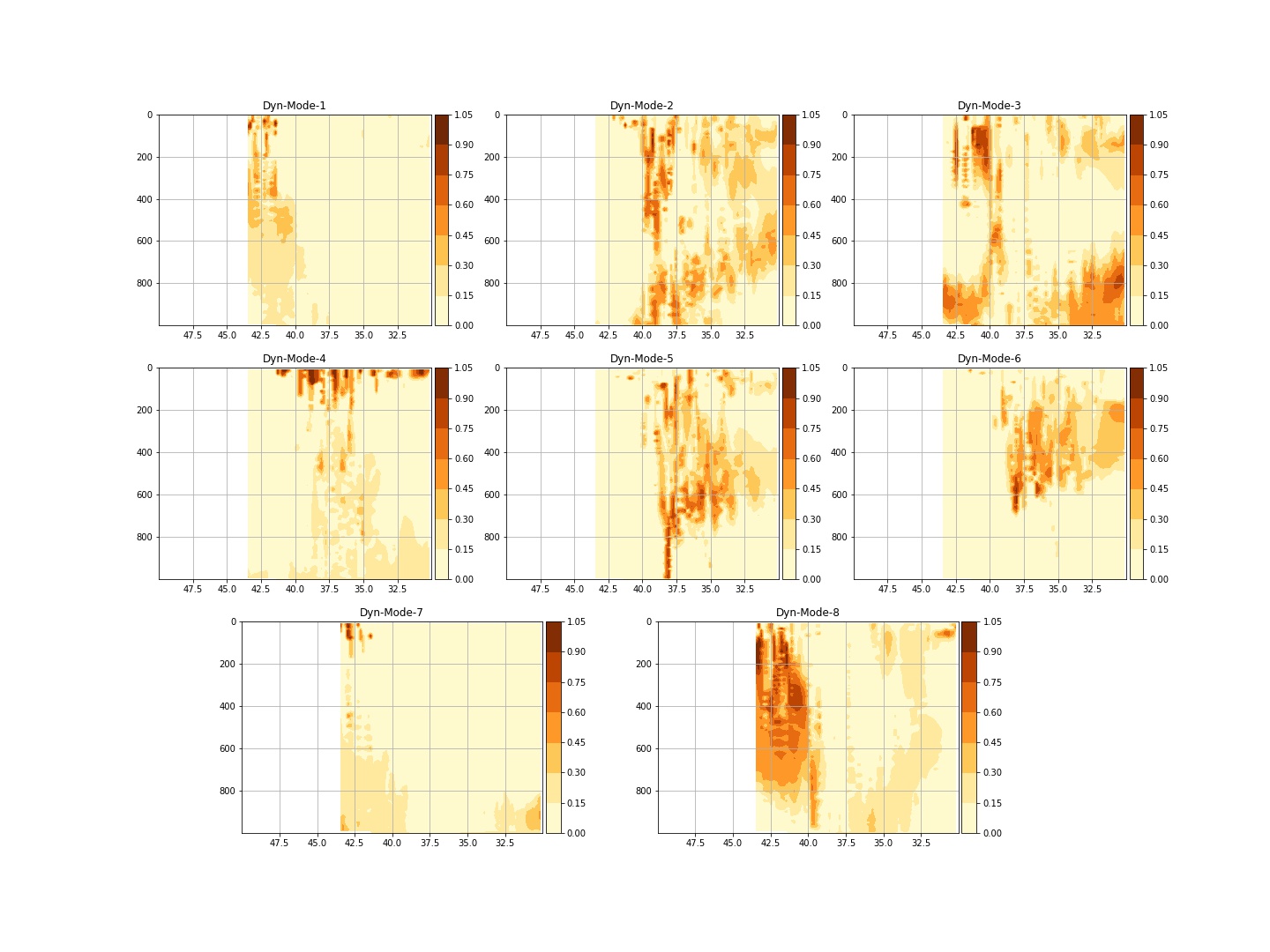}
     \caption{}
 \end{subfigure}
    \caption{Posterior maps for each dynamical mode (a) at the pressure 300, (b) at the longitude 60\si{\degree}W }
    \label{dynamic}
\end{figure}

Ocean current is one of the most concerned characteristics in oceanography. In the studied region, there are 3 dominant ocean currents, which are the Gulf Stream, the  Labrador and the North Atlantic. Each current has its own dynamic, and trajectory. We built posterior probability maps from all recorded data to analyse the current circulation in the region. The fig.~\ref{dynamic} (a) shows posterior maps for each dynamical mode in the region at the pressure 

We observed that a group dynamical modes could represent distinct circulation of each dominant current in the region. The $1^{st}$, $7^{th}$, and $8^{th}$ dynamical modes characterized a part of circulation of the Labrador current, which is a cold water region and a continuation of the Baffin Island Current and Flows southeastward from Hudson Strait (60\si{\degree}N) along the continental slope to the Tail of the Grand Banks (43\si{\degree}N). The $2^{nd}$, $3^{rd}$ and $5^{th}$ dynamical modes jointly characterize the Gulf Stream and North Atlantic currents, which are the warmer water regions. The North Atlantic current represents the bulk of the Gulf Stream continuation past its branch point, which is located at 60\si{\degree}W 60\si{\degree}N. The Gulf Stream region is located 40\si{\degree}N 50\si{\degree}W and one of the world’s most intensely studied current systems. To understand how the dynamical modes could deeply represent the Gulf Stream current system, we used a cut through the ocean at the longitude 60\si{\degree}W. In fig.~\ref{dynamic} (b) right  shows  posterior maps of dynamical modes through the cut.  The $2^{nd}$, $3^{rd}$ and $5^{th}$ dynamical modes of the maps deeply represent the Gulf Stream region in the latitude between 37.5\si{\degree}N and 40\si{\degree}N. \\

\subsubsection{Vertical structure of the seawater temperature and salinity in a turbulence region}
\begin{figure}
\raggedright
 \begin{subfigure}[b]{0.5\textwidth}
     \centering
     \includegraphics[width=\textwidth]{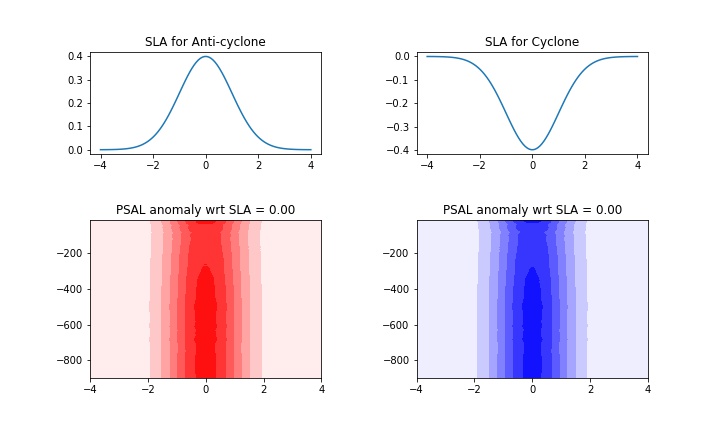}
%     \caption{}
 \end{subfigure}
 \begin{subfigure}[b]{0.5\textwidth}
     \centering
     \includegraphics[width=\textwidth]{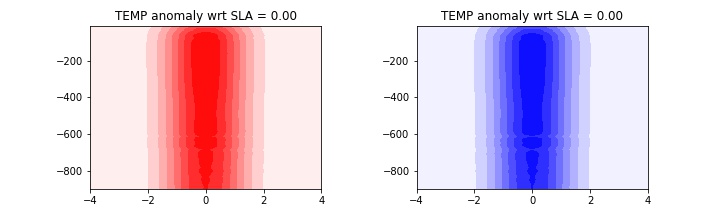}
%     \caption{}
 \end{subfigure}
    \caption{Vertical structure of the seawater temperature and salinity in cyclone and anticyclone turbulences}
    \label{vertstruct}
\end{figure}

When a turbulence passes through a local region of the ocean, it impacts on the seawater temperature and salinity in the region. To understand how the seawater temperature and salinity vertically distribute, we simulated two types of cyclones by varying values of the sea level anomaly. In fig.~\ref{vertstruct} represents these variations.\\

Once the posterior maps had been estimated for the local tiny region, we calculated the region’s temperature and salinity for a corresponding depth level and sea level anomaly. As we observed, the vertical structure is the same for the seawater temperature and salinity in two types of cyclones. In the anticyclone region, they have a warmer anomaly structure, and in the cyclone region, they posses a cooler anomaly structure.

\subsection{Second approach' results}
\subsubsection{Selection of the best model}
The architectures presented in the methodology section were tested. To do so, a cross-validation scheme was implemented by dividing the dataset in three subsets of equal size. Table \ref{tablecross} shows the results we obtained in each cross-validation for all different architectures. Cross-validation 1 means that the subset 1 from the dataset was used as validation data whereas subsets 2 and 3 were used for training purposes. Reported MSE always corresponds to validation data MSE.\\

It can be stated that the simple architecture is the one that describes ocean's temperature the best since it presents the lowest temperature's MSE. Nevertheless, the addition of an LSTM layer brings significant improvement in the description of salinity by considerably reducing MSE with respect to this variable.  Even though this architecture yields to higher temperature MSE when compared to the case in which the LSTM layer is absent, its performance is still better than that of all other considered architectures in terms of such quantity. For this reason, all the following results will be based on this architecture. \\

\begin{table}[t]
\centering
\resizebox{0.5\textwidth}{!}{
\begin{tabular}{c|c|c|c|c|c|c|l|l|}
\cline{2-9} & \multicolumn{2}{c|}{Cross-validation 1} & \multicolumn{2}{c|}{Cross-validation 2} & \multicolumn{2}{c|}{Cross-validation 3} & \multicolumn{2}{c|}{Average} \\ \hline
\multicolumn{1}{|c|}{Architecture}  & MSE T              & MSE S              & MSE T              & MSE S              & MSE T              & MSE S              & MSE T        & MSE S         \\ \hline
\multicolumn{1}{|c|}{Simple}        & 1.4775             & 0.07786            & 0.9280             & 0.04274            & 0.74611            & 0.06657            & \textbf{1.0505}       & 0.06239       \\ \hline
\multicolumn{1}{|c|}{Parallel}      & 1.8627             & 0.09192            & 0.8063             & 0.04001            & 0.72871            & 0.07628            & 1.1326       & 0.06940       \\ \hline
\multicolumn{1}{|c|}{Cascade}       & 1.9103             & 0.08647            & 0.8531             & 0.06497            & 0.9509             & 0.1024             & 1.2378       & 0.08461       \\ \hline
\multicolumn{1}{|c|}{Junction}      & 1.7702             & 0.1071             & 0.7860             & 0.04239            & 0.8632             & 0.07734            & 1.1398       & 0.07561       \\ \hline
\multicolumn{1}{|c|}{Simple + LSTM} & 1.5629             & 0.03877            & 0.8638             & 0.01799            & 0.8175             & 0.03518            & 1.0814       & \textbf{0.03065}       \\ \hline
\end{tabular}}
\caption{Cross-validation results. MSE for validation data regarding both temperature and salinity is shown for every considered architecture; the models presenting the best average results are highlighted in bold.}
\label{tablecross}
\end{table}

\subsubsection{Sea surface temperature prediction}
\begin{figure}
\raggedright
\centering
 \begin{subfigure}[b]{0.4\textwidth}
     \centering
     \includegraphics[width=\textwidth]{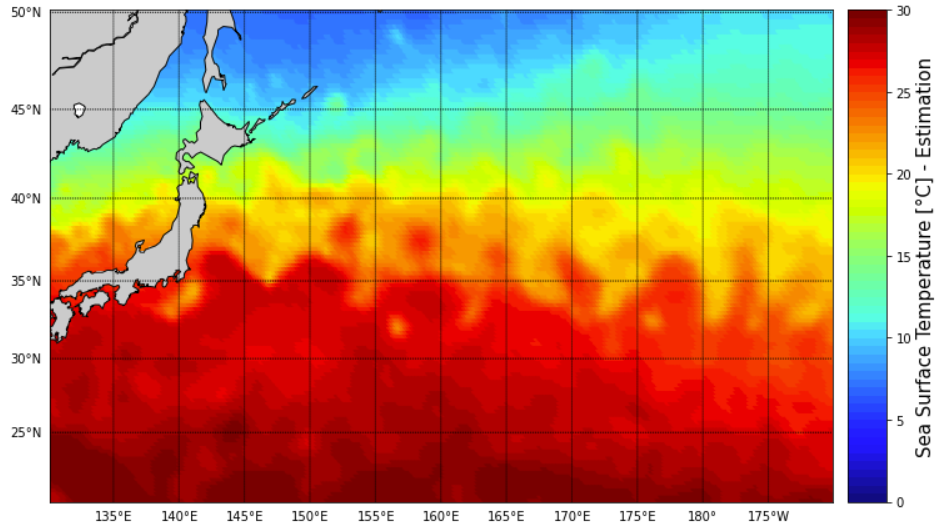}
%     \caption{}
 \end{subfigure}
 \begin{subfigure}[b]{0.4\textwidth}
     \centering
     \includegraphics[width=\textwidth]{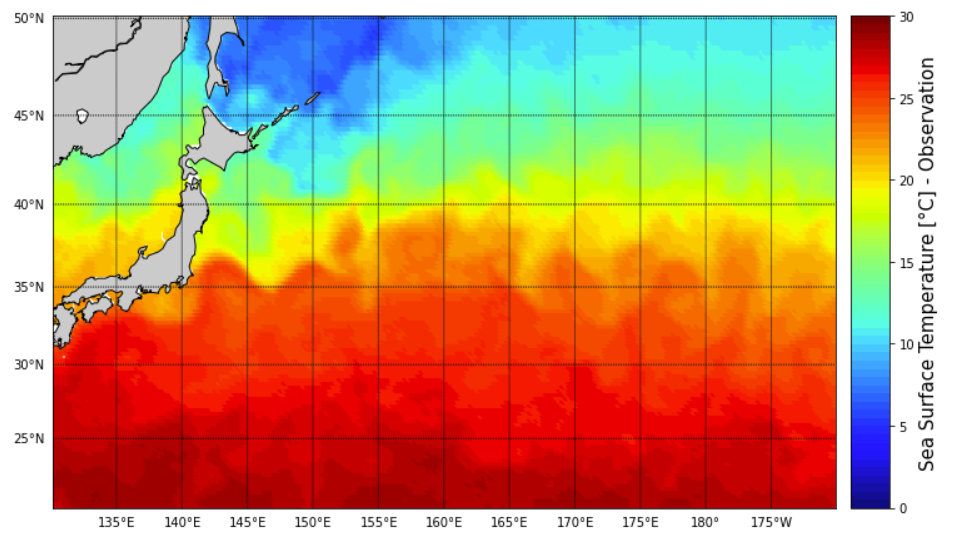}
%     \caption{}
 \end{subfigure}
 \begin{subfigure}[b]{0.4\textwidth}
     \centering
     \includegraphics[width=\textwidth]{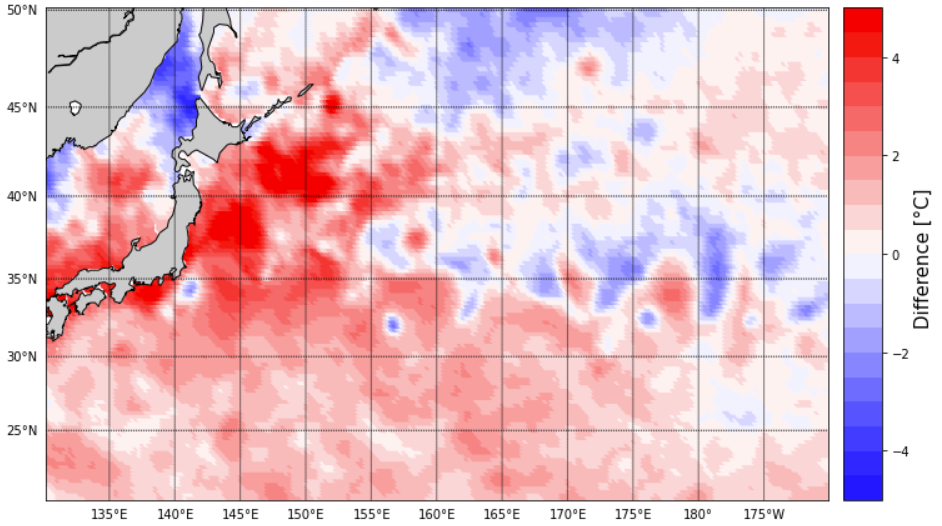}
%     \caption{}
 \end{subfigure}
    \caption{SST at Kuroshio's current extension (15/10/2015)}
    \label{map1}
\end{figure}

\begin{figure}
\raggedright
\centering
 \begin{subfigure}[b]{0.4\textwidth}
     \centering
     \includegraphics[width=\textwidth]{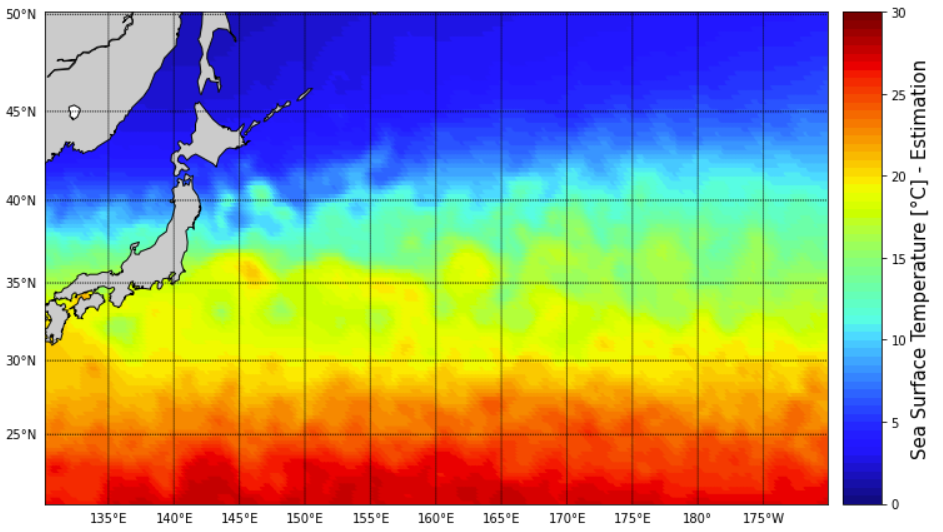}
%     \caption{}
 \end{subfigure}
 \begin{subfigure}[b]{0.4\textwidth}
     \centering
     \includegraphics[width=\textwidth]{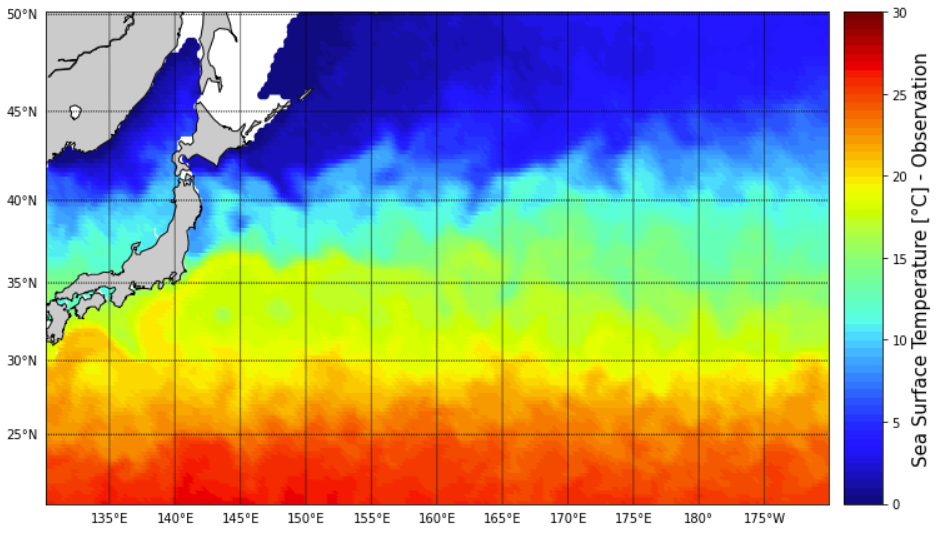}
%     \caption{}
 \end{subfigure}
 \begin{subfigure}[b]{0.4\textwidth}
     \centering
     \includegraphics[width=\textwidth]{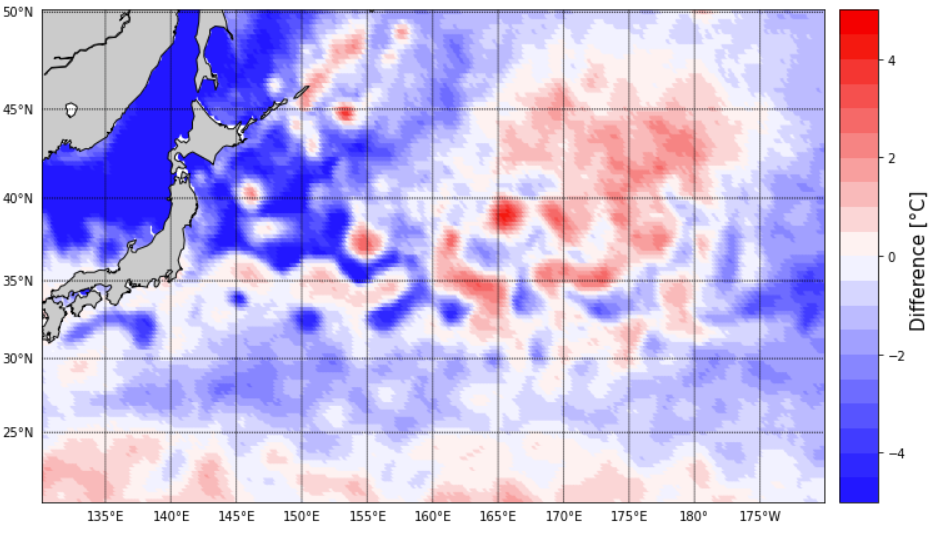}
%     \caption{}
 \end{subfigure}
    \caption{SST at Kuroshio's current extension (15/03/2020)}
    \label{map2}
\end{figure}

The model's ability to predict sea surface temperature was tested. Figure \ref{map1} shows average sea surface temperature at the Kuroshio's current extension on October 15, 2015. Figure \ref{map2} shows the same quantity on March 15, 2020. The observations our SST predictions are being compared to were extracted from Copernicus database, as well as the sla information necessary for our model calculation \cite{CMEMS2020}, \cite{CMEMS2020a}, \cite{CMEMS2020b}. \\

It can be stated that the model allows describing sufficiently well the sea surface temperature, with maximum absolute differences of about 5$^\circ$C near the coasts, where less data for training are available. However, the model doesn't manage to capture all temperature details and certainly further work can be done to better tune its parameters and therefore obtain better performances out of it. \\

\subsubsection{Temperature and salinity profiles}
\begin{figure}
\raggedright
\centering
 \begin{subfigure}[b]{0.4\textwidth}
     \centering
     \includegraphics[width=\textwidth]{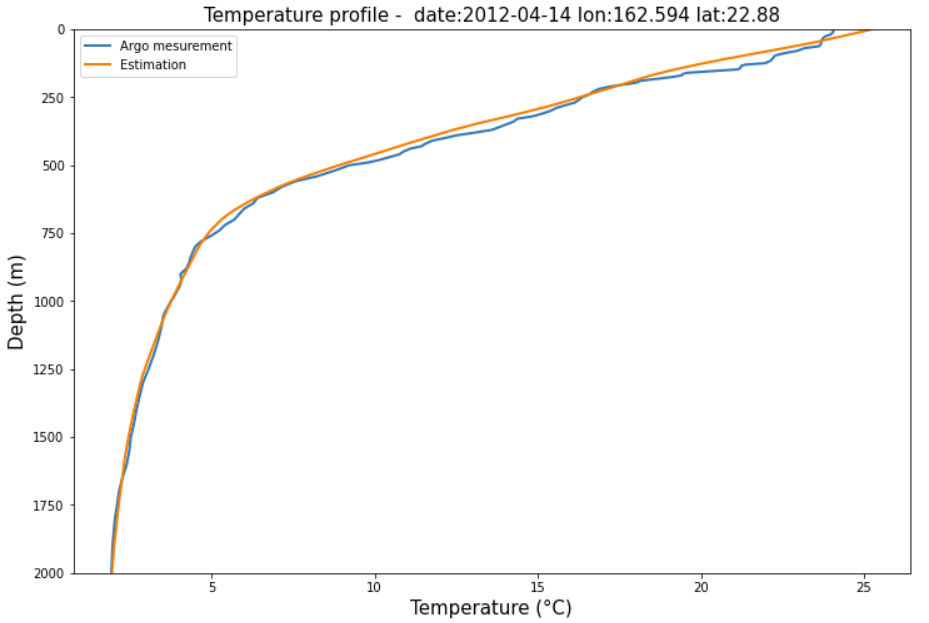}
     \caption{Temperature profile}
 \end{subfigure}
 \begin{subfigure}[b]{0.4\textwidth}
     \centering
     \includegraphics[width=\textwidth]{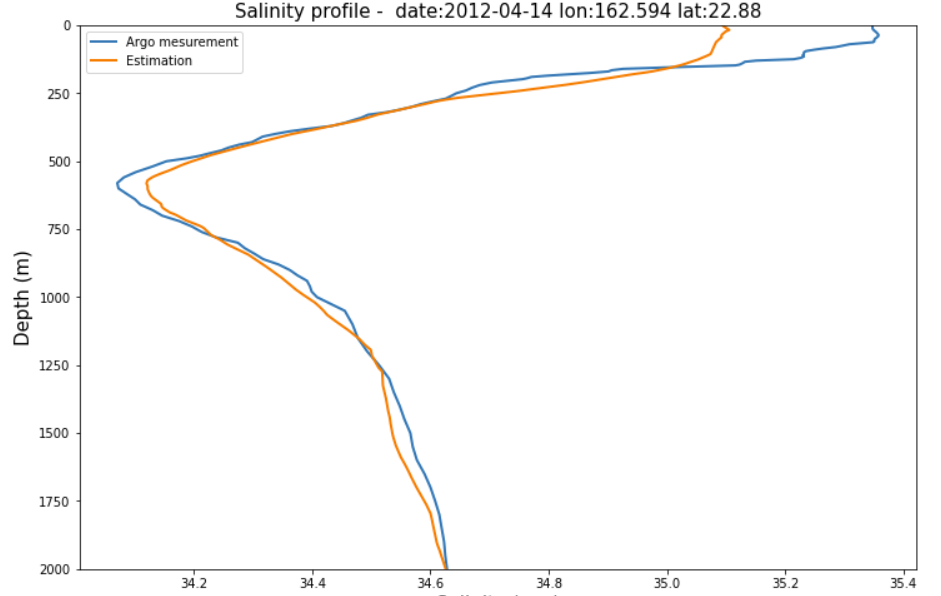}
  \caption{Salinity profile}
 \end{subfigure}
    \caption{Temperature and salinity's profile at long =162.594, lat=22.88 (162.594)}
    \label{profile1}
\end{figure}

\begin{figure}
\raggedright
\centering
 \begin{subfigure}[b]{0.4\textwidth}
     \centering
     \includegraphics[width=\textwidth]{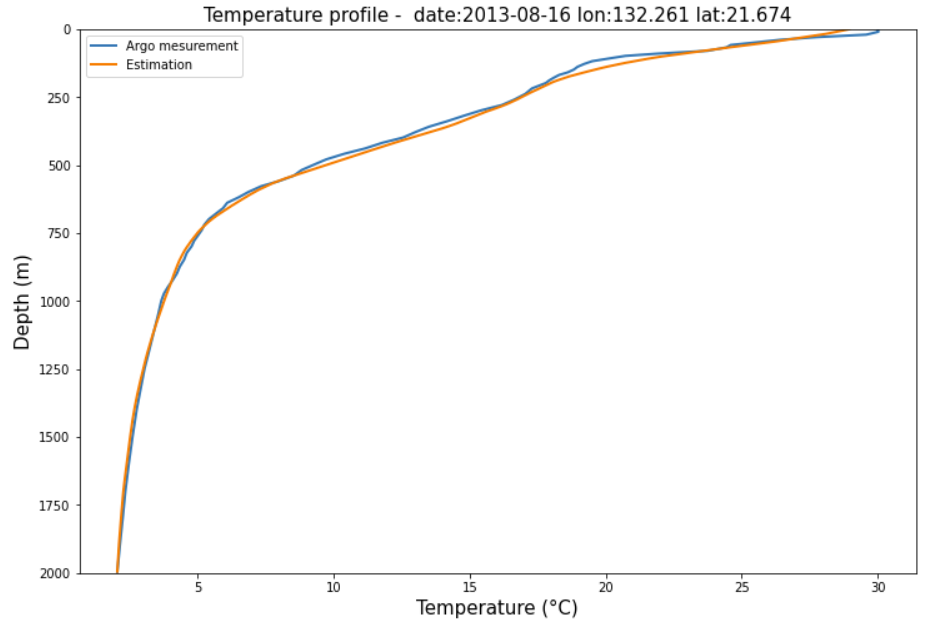}
     \caption{Temperature profile}
 \end{subfigure}
 \begin{subfigure}[b]{0.4\textwidth}
     \centering
     \includegraphics[width=\textwidth]{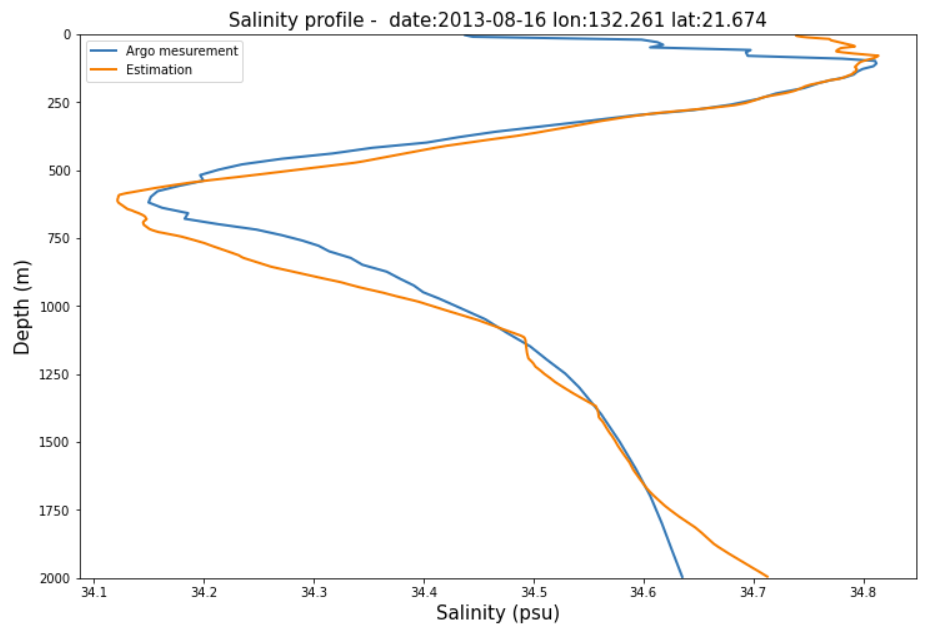}
  \caption{Salinity profile}
 \end{subfigure}
    \caption{Temperature and salinity's profile at long =132.261, lat=21.674 (16/08/2013)}
    \label{profile2}
\end{figure}

It is possible to predict temperature and salinity profiles at different geographical points for different dates using the model we trained. Some examples of such profiles are shown in figures \ref{profile1} and \ref{profile2}. It can be seen that the model allows accurately reproducing ocean's temperature by capturing most of its variability. Errors become lower as the depth increases, which shows that our model is less performing near and at the ocean's surface. Regarding salinity, our model does follow this variable's tendency but lacks of precision and doesn't manage to accurately predict its values. Further work may imply modifications to the current model to improve its performance in terms of salinity prediction.\\

\subsubsection{Vertical  structure  of  the  seawater  temperature  and salinity   in   a   turbulence   region}
To verify how our model manages to explain vertical distribution changes in temperature and salinity when a turbulence passes through a local region of the ocean, simulations with cyclone and anticyclone turbulence were performed. \\

\begin{figure}
     \centering
     \includegraphics[width=0.5\textwidth]{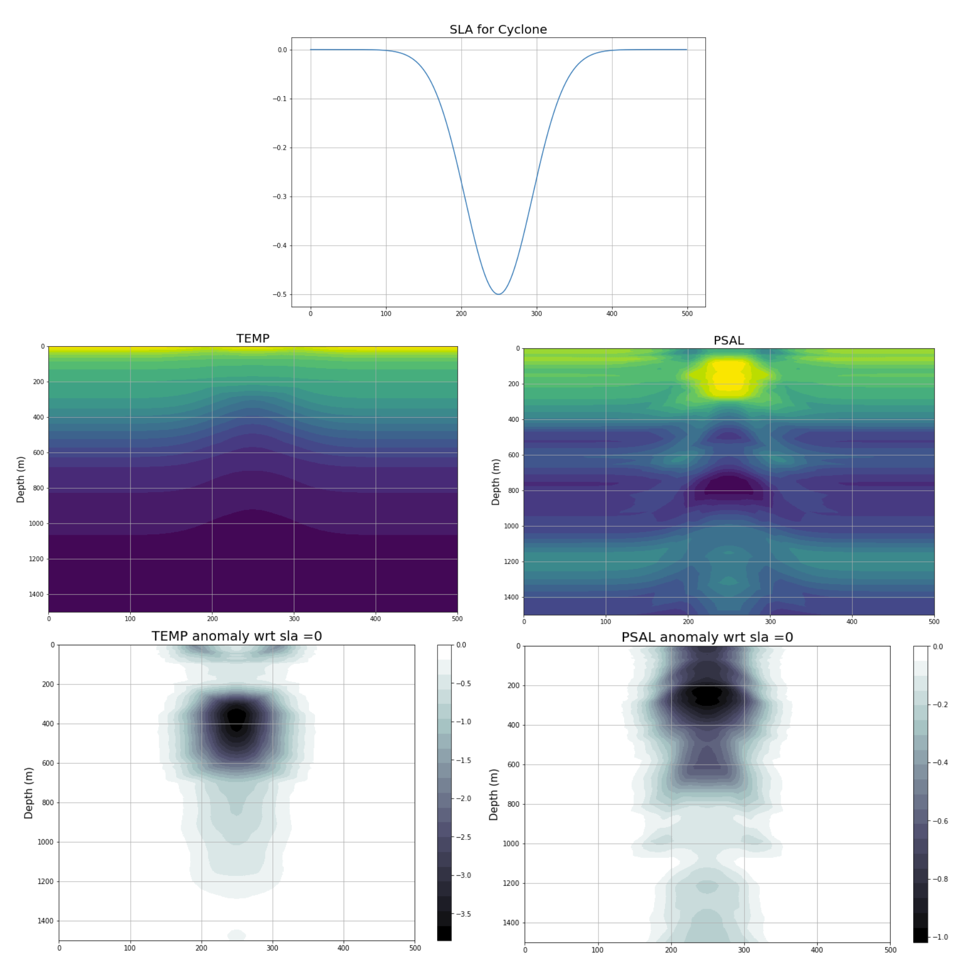}
    \caption{Temperature and salinity profiles in case of cyclone}
    \label{cyclone}
\end{figure}

\begin{figure}
     \centering
     \includegraphics[width=0.5\textwidth]{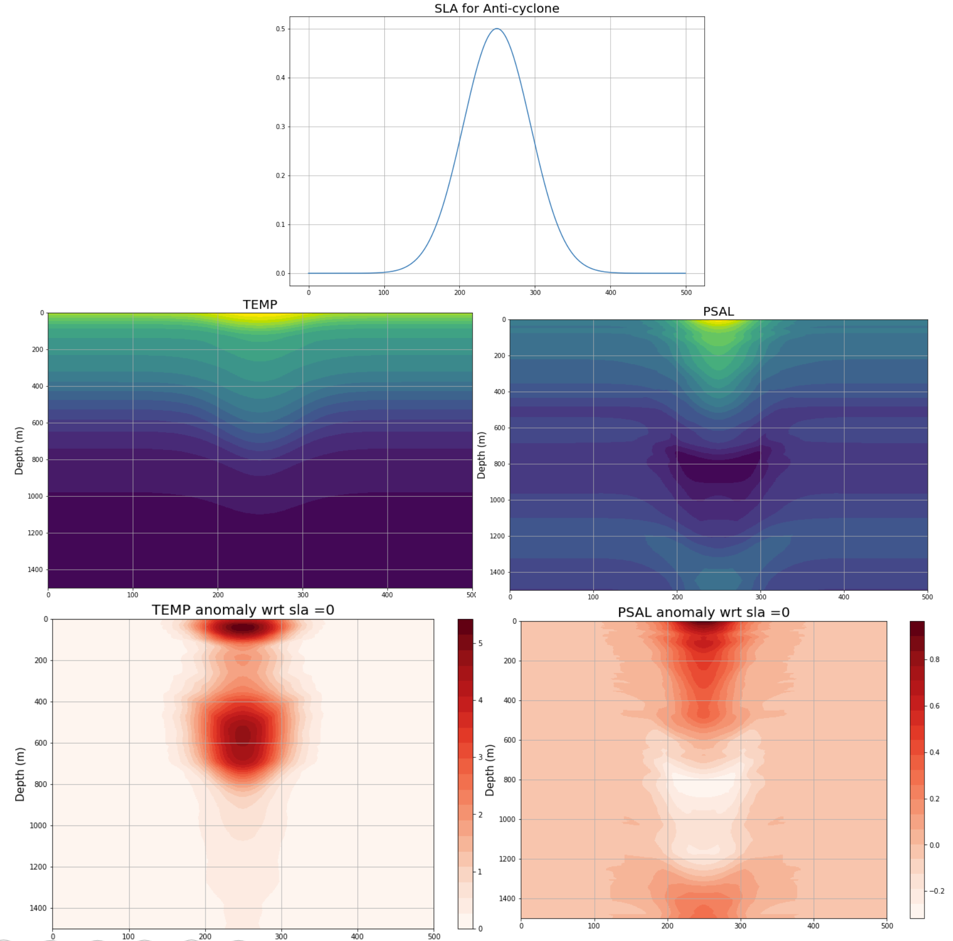}
    \caption{Temperature and salinity profiles in case of anticyclone}
    \label{anticylclone}
\end{figure}

Results are presented in figures \ref{cyclone} and \ref{anticylclone}, where the model's prediction in terms of temperature and salinity variations in turbulent situations can be seen. A warmer anomaly is detected in case of anticyclone, whereas a cooler one is detected in case of cyclone. \\ Salinity doesn't behave as temperature, which differs from the results regarding the statistical model based on latent regression as well as those presented in \cite{Tokunaga2019}. As it was stated above, the model based on neural networks doesn't manage to capture salinity's variability accurately, which may explain this difference. Indeed, simulations suggest that a cyclone may reduce salinity at deeper zones of the ocean more than it would in certain zones which are closer to the surface, which is unlikely to happen. \\

\subsubsection{Temporal variations at moorings' locations}
Temporal behavior of temperature and salinity using our model were performed at the KEO Mooring location. In site measurements were extracted from Ocean Climate Stations database \cite{NOAA2020}. At near-surface depths, it is  possible to compare our model to data extracted from Copernicus database, which also provides us with information about sla at such location. In addition, our model is also compared to ISAS-15, a model that has been used to predict temperature and salinity seasonality at KEO  and has been trained with data from Argo database \cite{Kolodziejczyk2017}. \\

\begin{figure}[h]
\raggedright
\centering
 \begin{subfigure}[b]{0.4\textwidth}
     \centering
     \includegraphics[width=\textwidth]{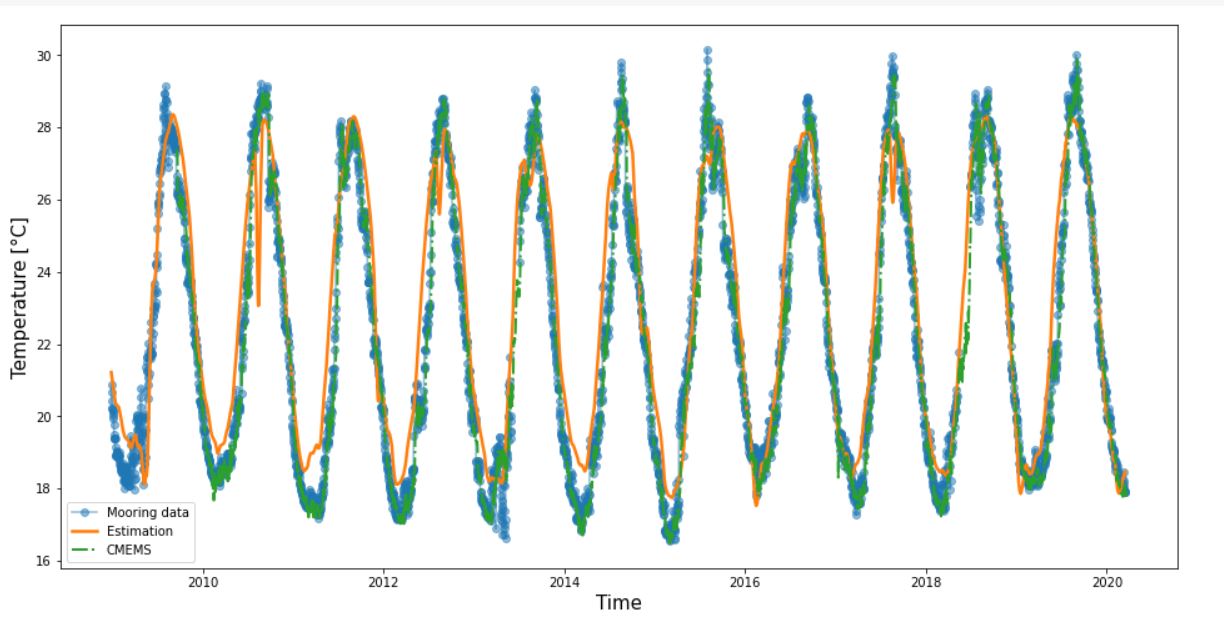}
     \caption{Surface time series}
 \end{subfigure}
 \begin{subfigure}[b]{0.4\textwidth}
     \centering
     \includegraphics[width=\textwidth]{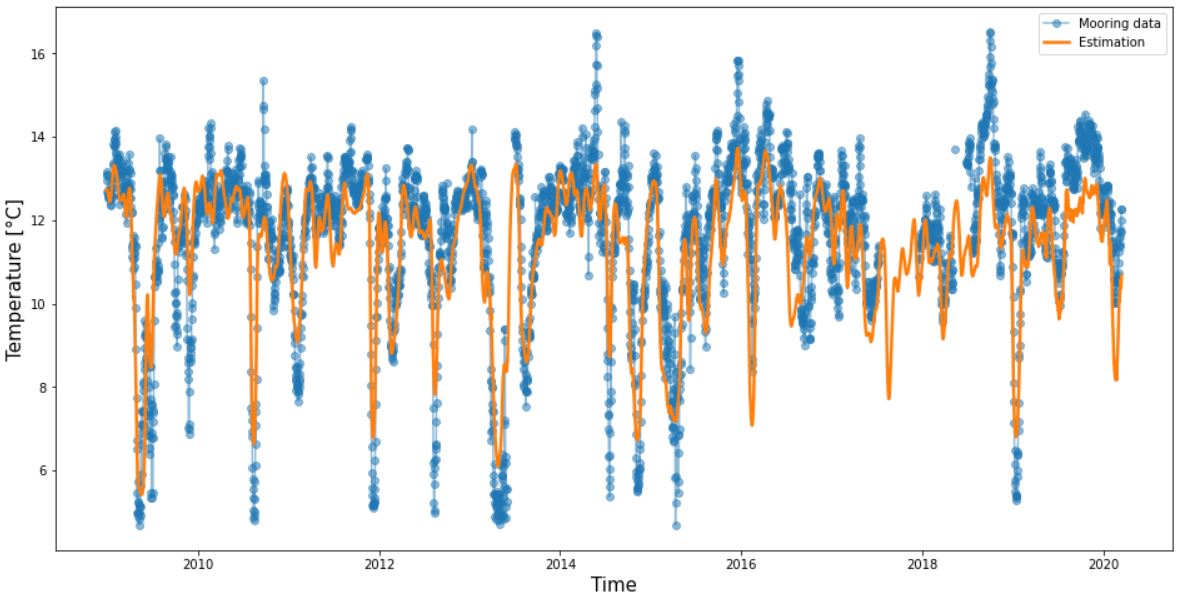}
     \caption{Time series at a depth of 525m}
 \end{subfigure}
    \caption{Temporal variation of the temperature at the surface and at a 525m depth.}
    \label{temporal}
\end{figure}

\begin{figure}[h]
\raggedright
\centering
 \begin{subfigure}[b]{0.4\textwidth}
     \centering
     \includegraphics[width=\textwidth]{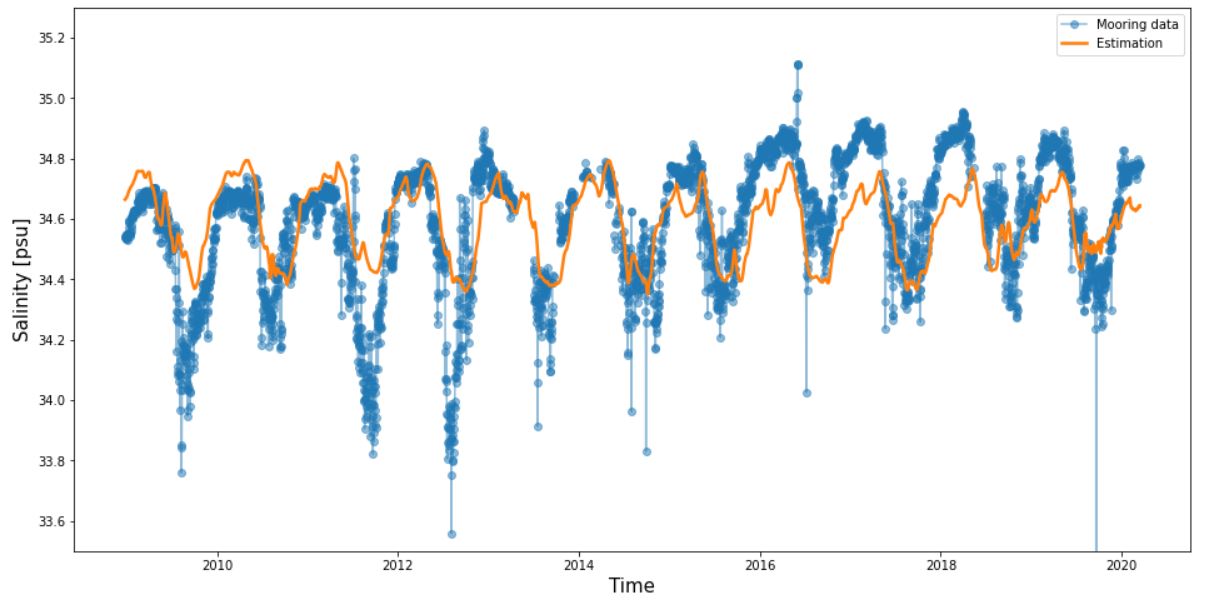}
     \caption{Surface time series}
 \end{subfigure}
 \begin{subfigure}[b]{0.4\textwidth}
     \centering
     \includegraphics[width=\textwidth]{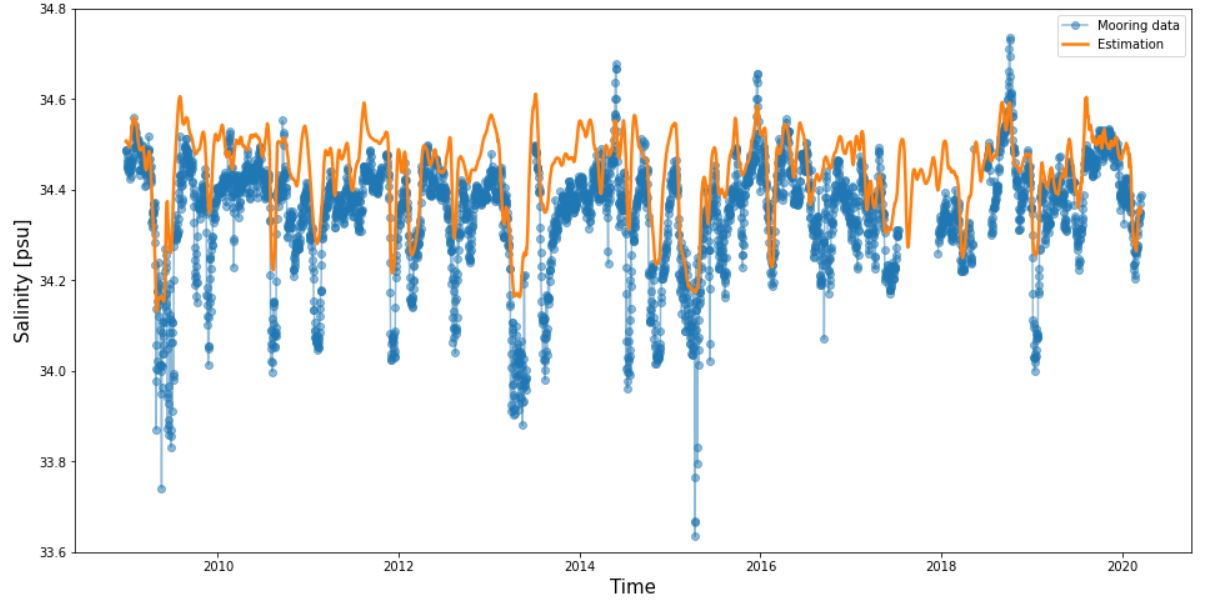}
     \caption{Time series at a depth of 525m}
 \end{subfigure}
    \caption{Temporal variation of salinity at the surface and at a 525m depth.}
    \label{temporal2}
\end{figure}

Figure \ref{temporal} shows temporal temperature variations between 2008 and 2020 at variable depths. It can be stated that we manage to predict the tendency of temperature in time, though some variations exist between our model and different available measurements. The curve corresponding to our model's results takes into account temperature values at midnight whereas most data we're comparing it to correspond to the daily mean, which may be a source of error explaining the observed differences.\\
Similar conclusions can be drawn from salinity variations presented at figure \ref{temporal2}, even though more representative errors exist. As stated before, our model doesn't accurately explain variability and salinity and further modifications are required to boost-up its performance with respect to this quantity.\\

\begin{figure}[h]
\raggedright
\centering
 \begin{subfigure}[b]{0.4\textwidth}
     \centering
     \includegraphics[width=\textwidth]{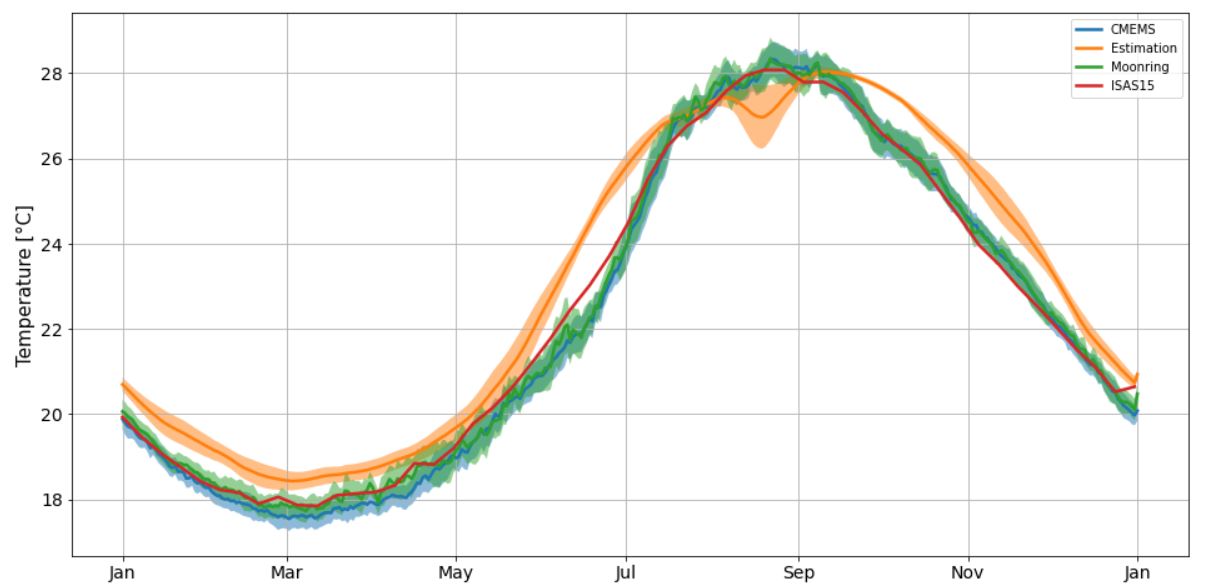}
     \caption{Temperature seasonality}
 \end{subfigure}
 \begin{subfigure}[b]{0.4\textwidth}
     \centering
     \includegraphics[width=\textwidth]{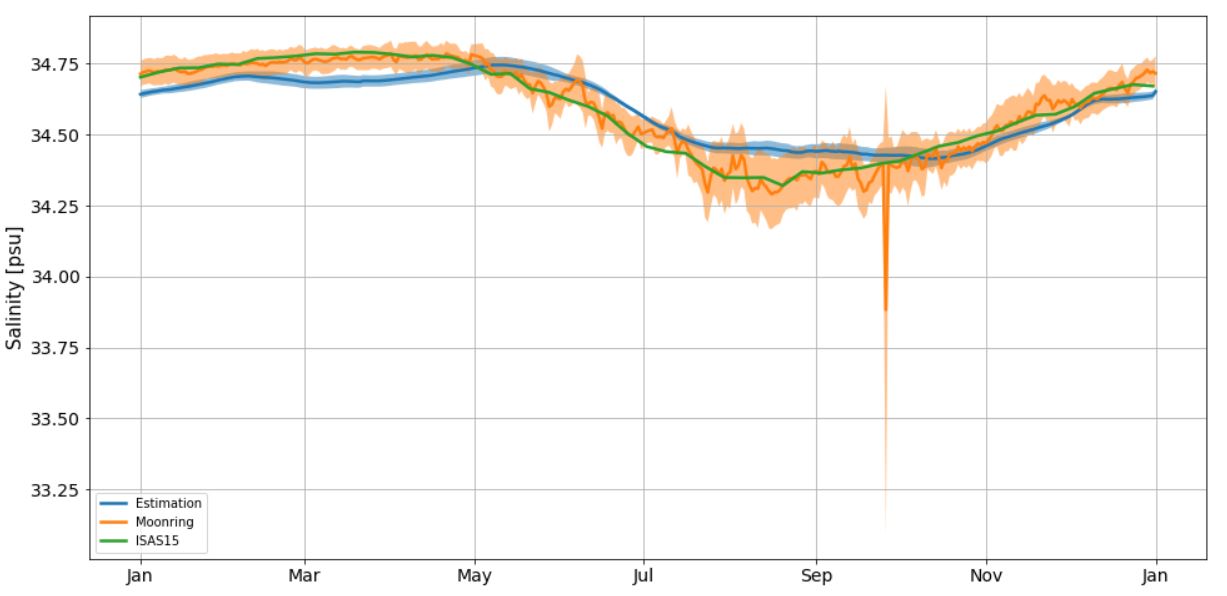}
     \caption{Salinity seasonality}
 \end{subfigure}
    \caption{Seasonality for temperature and salinity}
    \label{season}
\end{figure}

Time series calculated with our models were used to predict seasonality in temperature and salinity. Results are presented in figure \ref{season}. It is possible through our model to obtain a reasonable explanation of both variables seasonal tendencies, though it would be convenient to explore different possible modifications to improve its performance in terms of accuracy.

%=======================================================================================================
\section{Discussion \& Conclusions}

In our research problem, the Gaussian Mixture Model was used to discover and exploit the three dimensional structure of the sea water temperature and salinity in a part of the North Atlantic region. We worked on a notion of latent classes, which were dynamical modes to characterize the sea water temperature and salinity structure, which was also physically meaningful information that brought to us some insights to understand and to predict mesoscale turbulence movement in the ocean. 

The GMM method was used to interpolate the sea water temperature and salinity in the three dimensional ocean space, which yielded significantly low mean squared error in the climatology criteria. In some local regions, where the data collection was not continuous and interrupted,  had higher error than the others. The GMM method could be used to capture seasonal variation of the seawater temperature and salinity at specific local space in the ocean, which showed thoroughly these variations when it was located at the surface. Each dynamical mode characterized a specific structure of the region. In our analysis, we investigated 8 dynamical modes for describing  entirely the three dimensional structure of the part of the North Atlantic region. The 8 modes were used to guarantee a precise interpolation result of the seawater temperature and salinity, but it was a little bit harder to figure out the internal structure of the ocean, especially the ocean currents. The oceanic environment is also chaotic and changed by the impacts of mesoscale  turbulences. An analysis on the sea water temperature and salinity change, at an observation point,  under an impact of the cyclone and anticyclone turbulences were conducted. Predicting these changes required an estimation of posterior probabilities; the results showed accurate trends in the seawater temperature and salinity, but the accuracies are still limited. 

The Expectation-Maximization algorithm is a simple iterative method to find the maximum likelihood of a objective function. However, the optimization method is still sensitive to local maximum, and costly in terms of computation time for a large dataset. The BIC box plots in the figure.~\ref{boxplot} showed that the EM algorithm was approximately converged the same maximum likelihood solutions for different initialization. Moreover, with the model was not so complicated, which was the fuzzy regression based on linear ones, and the  the likelihood maximum procedure was simple, applied on a large dataset, we did not expect state-of-art results. But at least, through our experiences we could generally figure out some physical phenomena which we concern in the oceanography. \\

Regarding the neural network-based model, lots of hyperparameters must be tuned and they all influence in certain way the overall performance of the model (activation function, optimizer, goal function, batch size). In addition, the choice of the architecture as well as the network's number of parameters must take into account the complexity vs accuracy trade-off, which isn't necessarily simple to manage. This being said, there exists lots of degrees of freedom which make it particularly difficult to define a model and to conclude it does correspond to the best possible performances one can get from an ANN approach. \\

In our approach, most choices were made with the objective of boosting the network's accuracy with respect to validation data while keeping its structure as simple as possible. Different schemes were tested and the best one was retained in order to obtain all the results presented in this paper. It could be stated that the model allows predicting sufficiently well the temperature of the ocean: it manages to capture this quantity's variability at almost any depth, though it has better precision in deeper zones than it does in the surface. Nevertheless, the model doesn't accurately capture salinity's variability and even though it does follow this variable's tendency, significant errors persist, which make the model unusable for salinity prediction purposes since it risks leading to false conclusions. 
Our model is certainly imperfect and further work is necessary to improve it and get better performances from it, especially with respect to salinity prediction. However, this work shows the pertinence of using neural networks and the potential this kind of approach has to model complex dynamics such as those of the ocean in turbulent regions like the Kuroshio's current extension. 

%=======================================================================================================
\section{Appendix}
\begin{figure*}[t]
{ \centering
 %%%%
 \begin{subfigure}[b]{0.2\textwidth}
    \centering
     \includegraphics[width=1.1\textwidth]{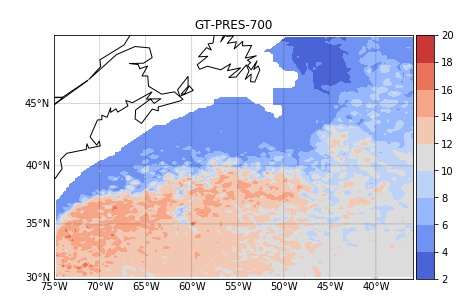}
 \end{subfigure}
\hspace{1em}%
 \begin{subfigure}[b]{0.2\textwidth}
    \centering
     \includegraphics[width=1.1\textwidth]{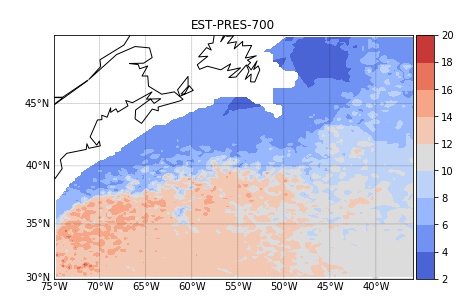}
 \end{subfigure}
\hspace{1em}%
 \begin{subfigure}[b]{0.2\textwidth}
    \centering
     \includegraphics[width=1.1\textwidth]{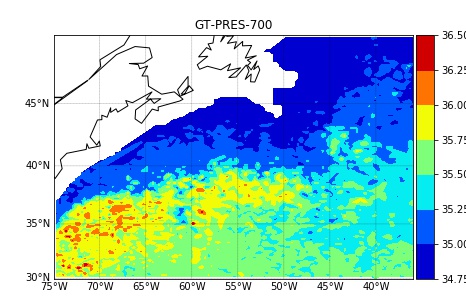}
 \end{subfigure}
\hspace{1em}%
 \begin{subfigure}[b]{0.2\textwidth}
    \centering
     \includegraphics[width=1.1\textwidth]{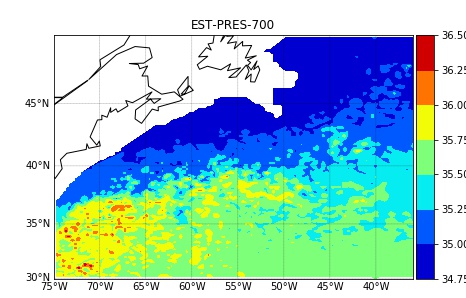}
 \end{subfigure}
 %%%%
  \begin{subfigure}[b]{0.2\textwidth}
    \centering
     \includegraphics[width=1.1\textwidth]{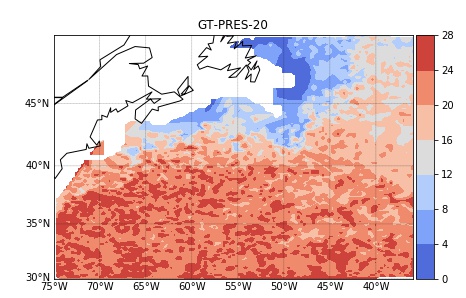}
 \end{subfigure}
\hspace{1em}%
 \begin{subfigure}[b]{0.2\textwidth}
    \centering
     \includegraphics[width=1.1\textwidth]{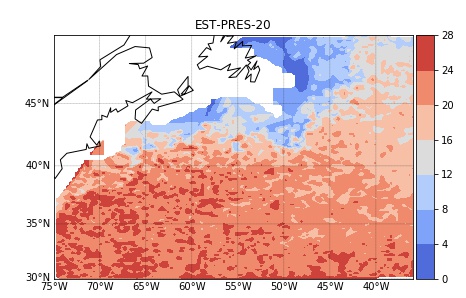}
 \end{subfigure}
\hspace{1em}%
 \begin{subfigure}[b]{0.2\textwidth}
    \centering
     \includegraphics[width=1.1\textwidth]{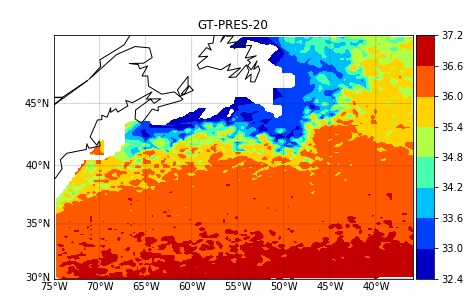}
 \end{subfigure}
\hspace{1em}%
 \begin{subfigure}[b]{0.2\textwidth}
    \centering
     \includegraphics[width=1.1\textwidth]{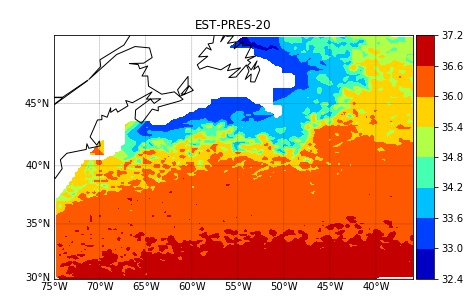}
 \end{subfigure}
%%%%
  \begin{subfigure}[b]{0.2\textwidth}
    \centering
     \includegraphics[width=1.1\textwidth]{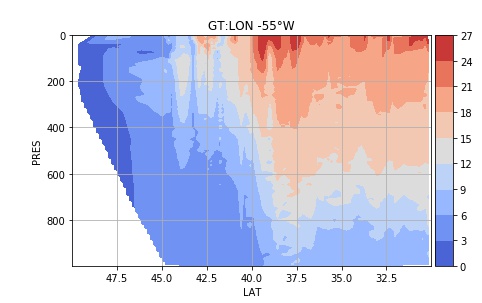}
 \end{subfigure}
\hspace{1em}%
 \begin{subfigure}[b]{0.2\textwidth}
    \centering
     \includegraphics[width=1.1\textwidth]{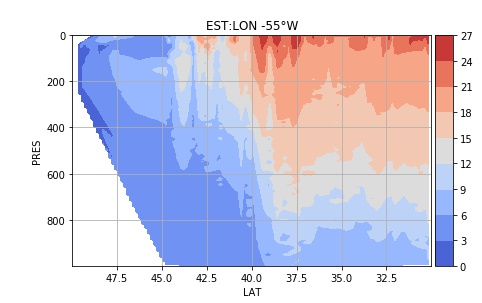}
 \end{subfigure}
\hspace{1em}%
 \begin{subfigure}[b]{0.2\textwidth}
    \centering
     \includegraphics[width=1.1\textwidth]{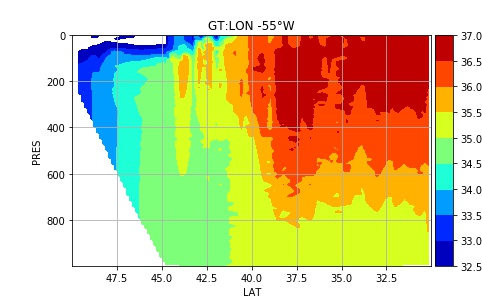}
 \end{subfigure}
\hspace{1em}%
 \begin{subfigure}[b]{0.2\textwidth}
    \centering
     \includegraphics[width=1.1\textwidth]{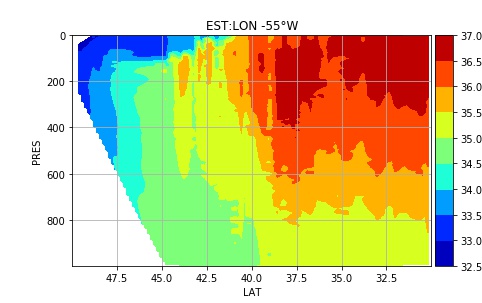}
 \end{subfigure}
%%%%
  \begin{subfigure}[b]{0.2\textwidth}
    \centering
     \includegraphics[width=1.1\textwidth]{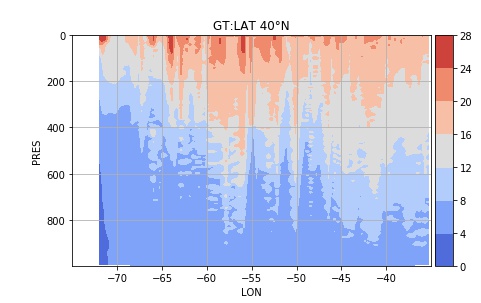}
     \caption{Ground truth Temperature}
 \end{subfigure}
\hspace{1em}%
 \begin{subfigure}[b]{0.2\textwidth}
    \centering
     \includegraphics[width=1.1\textwidth]{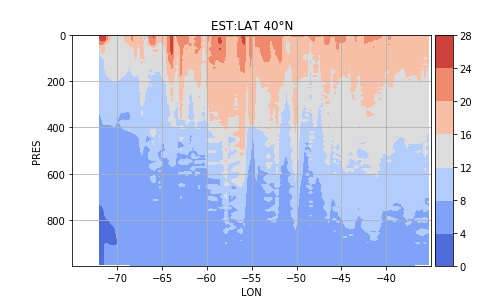}
     \caption{Estimated Temperature}
 \end{subfigure}
\hspace{1em}%
 \begin{subfigure}[b]{0.2\textwidth}
    \centering
     \includegraphics[width=1.1\textwidth]{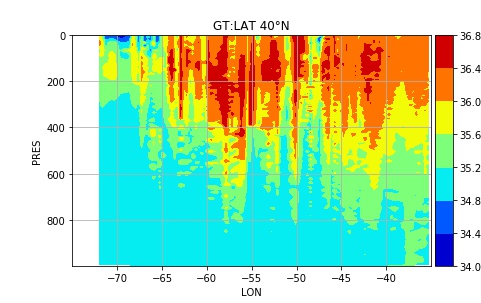}
     \caption{Ground truth Salinity}
 \end{subfigure}
\hspace{1em}%
 \begin{subfigure}[b]{0.2\textwidth}
    \centering
     \includegraphics[width=1.1\textwidth]{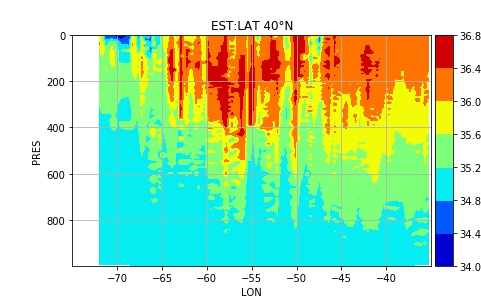}
     \caption{Estimated Salinity}
 \end{subfigure}

    \caption{Ground truth, estimated temperature and salinity (Gaussian Mixture Model) through a cut ($1^{st}$row) at the pressure 700, ($2^{nd}$row) at the pressure 20, ($3^{rd}$row) at the longitude 55\si{\degree}W, ($4^{th}$row) at the latitude 40\si{\degree}N}
    \label{gt-est}}
    
\end{figure*}

Figure \ref{gt-est} shows ground true estimated temperature and salinity (Gaussian Mixture Model) through a cut ($1^{st}$row) at the pressure 700, ($2^{nd}$row) at the pressure 20, ($3^{rd}$row) at the longitude 55\si{\degree}W, ($4^{th}$row) at the latitude 40\si{\degree}N. 

%=======================================================================================================
\bibliographystyle{ieeetr}
\bibliography{bibliography}

%=======================================================================================================
\section{Acknowledgement}
We would like to thank Assoc.Prof. Pierre TANDEO at the IMT Atlantique in Brest, and Research Scientist Guillaume MAZE at the IFREMER - French Institute for Ocean Science for their valuable time and expert advice, as well as their encouragement throughout the difficulty in this research project development. 
\clearpage
%=======================================================================================================
%Annexes

\end{document}